\documentclass{elsart}
\usepackage{natbib}
\usepackage{epsfig}
\newcommand{\Cv}{Cherenkov\ }
 
\def\lsim{\mathrel{\hbox{\rlap{\hbox{\lower4pt\hbox{$\sim$}}}
\hbox{$<$}}}}
 
\def\gsim{\mathrel{\hbox{\rlap{\hbox{\lower4pt\hbox{$\sim$}}}
\hbox{$>$}}}}

\begin{document}

\begin{frontmatter}
\title{VERITAS:
the Very Energetic Radiation Imaging Telescope Array System}
\author[sao]{T.C. Weekes}
\author[sao]{H. Badran}
\author[ou]{S.D. Biller}
\author[ul]{I. Bond}
\author[ul]{S. Bradbury}
\author[wu]{J. Buckley}
\author[isu]{D. Carter-Lewis}
\author[sao]{M. Catanese}
\author[sao]{S. Criswell}
\author[pu]{W. Cui}
\author[wu]{P. Dowkontt}
\author[gc]{C. Duke}
\author[nuid]{D.J. Fegan}
\author[pu]{J. Finley}
\author[uc]{L. Fortson}
\author[pu]{J. Gaidos}
\author[nuig]{G.H. Gillanders}
\author[sao]{J. Grindlay}
\author[isu]{T.A. Hall}
\author[sao]{K. Harris}
\author[ul]{A.M. Hillas}
\author[sao]{P. Kaaret}
\author[dpu]{M. Kertzman}
\author[uu]{D. Kieda}
\author[isu]{F. Krennrich}
\author[nuig]{M.J. Lang}
\author[isu]{S. LeBohec}
\author[pu]{R. Lessard}
\author[ul]{J. Lloyd-Evans}
\author[ul]{J. Knapp}
\author[nuid]{B. McKernan}
\author[uw]{J. McEnery}
\author[gmit]{P. Moriarty}
\author[uc]{D. Muller}
\author[ul]{P. Ogden}
\author[ucla]{R. Ong}
\author[isu]{D. Petry}
\author[nuid]{J. Quinn}
\author[ksu]{N.W. Reay}
\author[cit]{P.T. Reynolds}
\author[ul]{J. Rose}
\author[uu]{M. Salamon}
\author[pu]{G. Sembroski}
\author[ksu]{R. Sidwell}
\author[sao]{P. Slane}
\author[ksu]{N. Stanton}
\author[uc]{S.P. Swordy}
\author[uu]{V.V. Vassiliev}
\author[uc]{S.P. Wakely}
\address[sao]{Whipple Observatory, Harvard-Smithsonian Center for
Astrophysics,\\ P.O. Box 97, Amado, AZ 85645-0097 USA}
\address[ou]{Department of Physics, University of Oxford, OX1
3RH, U.K.}
\address[ul]{Department of Physics \& Astronomy, University of
Leeds, Leeds, LS2 9JT, U.K.}
\address[wu]{Department of Physics, Washington University,\\
Campus Box 1105, One Brookings Drive, St. Louis, MO 63130 USA}
\address[isu]{Department of Physics \& Astronomy, Iowa State
University,\\   Ames, IA, 50011, USA  }
\address[nuid]{Department of Experimental Physics,
National University of Ireland, Dublin,\\
Belfield, Dublin 4, Ireland}
\address[pu]{Department of Physics, Purdue University,\\
West Lafayette IN 47907-1396 USA}
\address[gc]{Department of Physics, Grinnell College, \\
 Grinnell, IA 50112 USA}
\address[nuig]{Department of Experimental Physics, National
University of Ireland,\\ Galway, Ireland}
\address[dpu]{Department of Physics \& Astronomy, DePauw
University,\\ Greencastle, IN 46135 USA}
\address[uu]{Department of Physics, University of Utah,\\
Salt Lake City, UT 84112 USA}
\address[uw]{Department of Physics, University of Wisconsin,\\
Madison, WI 53706 USA}
\address[gmit]{Galway-Mayo Institute of Technology, Galway,
Ireland}
\address[uc]{Enrico Fermi Institute, The University of Chicago,\\
5640 South Ellis Avenue, Chicago, IL 60637-1433 USA}
\address[ucla]{Department of Physics \& Astronomy, University
of California,\\
Los Angeles, CA 90024 USA}
\address[ksu]{Department of Physics, Kansas State University,\\
Manhattan, KS 66506-2601 USA}
\address[cit]{Cork Institute of Technology, Cork, Ireland}

\begin{abstract} 

The Very Energetic Radiation Imaging Telescope Array
System (VERITAS) represents an important step forward in the study of
extreme astrophysical processes in the universe.  It combines the
power of the atmospheric Cherenkov imaging technique using a large
optical reflector with the power of stereoscopic observatories
using arrays of separated
telescopes looking at the same shower.  The seven identical
telescopes in VERITAS, each of aperture 10 m,
 will be deployed in a filled hexagonal
pattern of side 80 m; each telescope will have a camera
consisting of 499 pixels with a field of view of 3.5$^\circ$. 
VERITAS will substantially increase the catalog of very high energy
(E$>$100\,GeV) $\gamma$-ray sources and greatly improve
measurements of established sources.
\end{abstract}


\end{frontmatter}

%%%%%%%%%  INTRO
%%%%%%%%%%%%%%%%%%%%%%%%%%%%%%%%%%%%%%%%%%%%%%%%%%%%%%%%%
\section{Introduction}
%FINAL
\subsection{Very High Energy Gamma-ray Astronomy}

Very High Energy (VHE) $\gamma$-rays are unique probes of the
Universe.
Unlike radiation at longer wavelengths, $\gamma$-rays are not
attenuated significantly within the Galaxy.  As such, they provide
an
unobstructed view through the plane of the Galaxy except for
regions
which contain high densities of low energy photons (or virtual
photons
in regions of high magnetic fields) where VHE $\gamma$-rays are
attenuated via photon-photon pair production.  The low photon
density requirement 
provides information about possible production sites near active
galactic nuclei (AGN) as does the low  magnetic field requirement
near pulsars. For $\gamma$-rays from
extragalactic sources, pair production with low energy photons can
be
used to probe intergalactic radiation fields.  VHE
$\gamma$-rays are at the end of the
electromagnetic
emission spectra of the objects which produce them, so the VHE
$\gamma$-ray emission features are often the most sensitive
probes of source
emission models. 

  The field of ground-based $\gamma$-ray astronomy has been
revolutionized by the development of the atmospheric Cherenkov
imaging
technique for the discovery and study of individual sources \cite
{Ong98,Aharonian97,Hoffman97,CataneseWeekes99}.
Although less than 1\% of the sky has been
surveyed by the imaging technique at energies of 300\,GeV
and above, thirteen sources
have
now been reported (eight with high significance) \cite{Weekes99}:
three pulsar-powered nebulae, six
BL Lacertae-type active galactic nuclei, three shell-type
supernova remnants, and one X-ray binary system.
These measurements have advanced our understanding of the origin of
cosmic rays, the nature of AGN jets, the density of the background
infrared (IR) radiation, and the magnetic fields in the shells and
nebulae of supernova remnants (SNR).

In this paper the technique and the imperative to progress to a next
generation observatory is first described (Section 2). Then the design
of the Very
Energetic Radiation Imaging Telescope Array System (VERITAS), one
such observatory, is presented (Section 3). The
technical aspects of VERITAS are described in Section 4. The
scientific potential of next generation ground-based telescopes is
outlined in Section 5. Finally, in Section 6 the projected
sensitivity of VERITAS is compared with other observatories.

\section{The Atmospheric Cherenkov Imaging Technique}

\subsection{Imaging Technique}

Atmospheric Cherenkov $\gamma$-ray telescopes have an inherent
advantage over space-based $\gamma$-ray
telescopes because of their very large collection areas 
($>$40,000\,m$^2$ compared with $<$1\,m$^2$). However, since no
anti-coincidence shield can be used in air shower detectors, 
it is necessary to exploit
the differences between $\gamma$-ray showers and hadronic
showers 
to reject the large background of hadronic cosmic rays. 
Atmospheric
Cherenkov imaging was proposed in 1977 for this purpose
\cite{Weekes77}. The basis of the technique is the use of an array
of photomultiplier tubes (PMTs) in the focal plane of a large
optical
reflector to record the Cherenkov image of an air shower.
The geometry of the imaging technique is illustrated in Figure 1.
%Figure~\ref{shower}.

[INSERT FIGURE 1]

Since the first experiments, it has been known that there are basic
differences in the appearance of the Cherenkov light
image of a $\gamma$-ray shower and that of a typical background
hadron shower. There is a physical difference in the shower
development due to the smaller transverse momentum of
electromagnetic interactions
compared with hadronic interactions. This difference
 means that the Cherenkov radiating
particles in the $\gamma$-ray shower are, on average, closer to the
direction of the primary. In addition there is no penetrating
component in
electromagnetic showers, so the local contribution from individual
particles to the light is small and the fluctuations in the shower
image are less. Because the $\gamma$-ray images are better defined,
it is easier to characterize their arrival direction and hence to
discriminate them against the isotropic background of hadron
showers.

The power of the imaging technique was first demonstrated using an
 array of photomultiplier tubes (PMTs) in the focal plane of the
 Whipple Observatory 10\,m optical reflector \cite{Cawley90}; the
detection of the Crab Nebula was the first major success of the technique
 \cite{Weekes89}
\cite{Vacanti91}.  The background was rejected with 99.7\%
efficiency and the source location was determined with an accuracy
of 0.05$^\circ$.
The Whipple camera design has since been duplicated at a number of
observatories 
\cite{Barrau98,Hara93,Konopelko99b,Aiso97,
Chadwick99,Vladimirsky89,Nikolsky89,tactic97} 
where it has been successfully used
to extend the catalog of detected TeV sources.

\subsection{Imaging with Arrays}

The atmospheric Cherenkov imaging technique can be significantly
improved by the use of multiple telescopes with separations of the
same order as the lateral spread of the light from the shower.
  The technique has
been effectively demonstrated using pairs of telescopes
\cite{Krennrich95} and arrays \cite{Konopelko99}. 
The reduction in energy threshold is less than that achieved by
combining all the mirror area into one telescope; however 
the distribution of mirror area into multiple
telescopes gives many images
 of the same shower which offer the following advantages:

\begin{itemize}
\item Improved hadron discrimination from multiple image
characterization.
\item Elimination of local muon background.
\item Improved energy resolution from multiple measurements, shower
axis location and determination of shower maximum.
\item Improved angular resolution. 
\end{itemize}

\subsection{Next Generation Telescopes}

Advances made by the present generation of imaging
telescopes justify the construction of an
array of large imaging telescopes with the following properties:

\vspace*{-0.1in}
\begin{itemize}
\itemsep=0pt
\parsep=0pt
\parskip=0pt

\item {\it Large Effective Area}: $>$0.1\,km$^2$ to provide 
sensitive measurements of short variability time-scales.

\item {\it Better Flux Sensitivity}: detection of sources which
emit $\gamma$-rays at levels of 0.5\% of the Crab Nebula flux at
energies of 300\,GeV in 50 hours of observation.

\item {\it Reduced Energy Threshold}:  an effective energy
threshold $<$100\,GeV with significant sensitivity at
50\,GeV.

\item {\it Improved Energy Resolution}: an RMS spectral resolution
of $\Delta$E/E $<$ 0.10 - 0.15 for an individual shower over a broad
energy range (E $>$ 300 GeV). 

\item {\it Increased Angular Resolution}: $<$0.05$^\circ$ for
individual showers and source location better than 0.005$^\circ$
($>$ 100 photons).

\item  {\it Large Field of View:} at least 3$^\circ$
diameter as used in many current atmospheric \Cv imaging
telescopes.

\end{itemize}

\vspace*{-0.1in}

All of these objectives can be
achieved by VERITAS, an array of large imaging
telescopes which are improved versions of
the existing Whipple 10\,m imaging telescope
\cite{lewis90}. 
The seven telescopes in VERITAS will be identical and
will be deployed as shown in Figure 2.
%Figure~\ref{athena-fig}. 
Six telescopes
will be located at the corners of a hexagon of side 80\,m and one
will
be located at the center. The telescopes will each have a camera
consisting of 499 pixels with a field of view of 3.5$^{\circ}$
diameter.  

[INSERT FIGURE 2]
  
The VERITAS concept was first described in 1985 prior to the
definite detection of any TeV sources \cite{Weekes85}. The first
practical demonstration of the power of an array of imaging
telescopes came with the completion of HEGRA, an array of five 3.5m
aperture telescopes \cite{Konopelko99}.

%%%%% Design Studies (Array Performance?)
%%%%%%%%%%%%%%%%%%%%%%%%%%%%%%%%%%%%

\section{VERITAS Design}
%FINAL
The performance characteristics required of VERITAS are derived
from its scientific goals which are discussed in Section 5.
 VERITAS was designed to optimize
observations of a variety of sources, each
of which may require a different subset of capabilities (e.g.,
continuous monitoring, large field of view, good flux
sensitivity, large collection area, accurate angular resolution,
low
energy threshold, accurate energy resolution, prompt response, and
broad energy coverage).  With these features, VERITAS can observe
known
and anticipated sources in several observing modes.

Extensive simulation studies have been used to determine the
optimum
configuration for VERITAS and to characterize its performance;
these
are described in detail elsewhere \cite{Vassiliev98,
Vassiliev99,weekesveritas}.  The
VERITAS design is optimized for maximum sensitivity to point
sources
in the energy range 100\,GeV - 10\,TeV, but with significant
sensitivity in the range 50\,GeV - 100\,GeV and from 10\,TeV -
50\,TeV.

\subsection{Choice of VERITAS parameters}

To determine the optimum array configuration, several parameters
were varied: the number of telescopes, the spacing between the
telescopes, the focal length of the telescope, the aperture of the
telescopes, and the field of view (FOV) 
of the cameras.  The number of
channels in the camera was fixed before the optimization for cost
reasons.  Thus, the investigations of different camera fields of
view
are simultaneously investigations of changes in the spacing of the
PMTs.  In the medium energy range of 200\,GeV to 1\,TeV, the
sensitivity of the array to point sources is not significantly
affected by small changes in the array's characteristics. 
The
choice of telescope parameters was driven mostly by their effects
on the performance of VERITAS at the low and high ends of the
sensitive energy range.

\subsubsection{Number of Telescopes}
At a minimum, three telescopes are necessary to utilize the
stereoscopic imaging technique.  Two telescopes can be used for
bright
images, but faint cascades can only be accurately reconstructed
with
images in three telescopes because the individual image axes are
not
well-defined. There are three primary reasons for choosing a seven
telescope array over one with three to four telescopes with similar
characteristics.

1. Low Energies.

Although high sensitivity measurements in the energy region above
130\,GeV will be accessible with a smaller array of 10\,m
telescopes
(e.g., 3 or 4 telescopes), the energy range from 50\,GeV to
130\,GeV is only accessible with the full array of seven
telescopes.
Some cosmic sources (e.g., $\gamma$-ray pulsars or possibly
$\gamma$-ray bursts) may be detectable {\it only} in the lowest
energy
range.
However, energy flux 
measurements on {\it all} sources will benefit from
having well-defined spectral measurements over a wide range of
energies. The differential flux sensitivity is defined as the flux
level in an energy interval of 1/4 of a decade over which an
energy flux measurement can be made with less than 20\% error
in 5 hours of observation.  This is the meaningful level
for astrophysical
measurements. The flux sensitivity for the full array is shown in
Figure 3 
%Figure~\ref{fluxsensitivity} 
where it is contrasted with the flux
sensitivity for a three telescope array (sub-array). For reference, the 
measured spectrum of Markarian 421 \cite{krennrich99} observed during a
high state of emission is also shown.

[INSERT FIGURE 3]

2. Versatility

The observing program of VERITAS \cite{weekesveritas} 
can only be accomplished in a
timely
fashion if the array can be divided in two or more sub-arrays for
simultaneous monitoring of different regions of the sky. A
sub-array
of three or four telescopes is very powerful and performs almost as
well as the full array for the detection of new sources and
measurement of established sources above $\sim$250\,GeV
%(Figure~\ref{fluxsensitivity})
 (although only the full array will reach the lowest energies). 
Above $\sim$250\,GeV, a sub-array
also has angular resolution comparable to the full array and
provides
excellent performance for spectral studies.  Some scientific
studies,
such as the long-term variability of AGN, can {\it only} be
accomplished with a sub-array dedicated to monitoring one
particular
object for extensive time periods.  It is anticipated that the
array
will be operated as two independent sub-arrays at least 50\% of the
time.

3. Flexibility

Experience has shown that the full potential of an atmospheric
Cherenkov telescope is only realized {\it after} the detection of
$\gamma$-ray sources. Therefore, it is anticipated that, with
operating
experience, significant improvements will be made in the achievable
flux sensitivity of VERITAS by varying the observing techniques,
the telescope hardware, and the data analysis algorithms. The
multi-telescope configuration of
VERITAS presents the maximum adaptability in this regard.

\subsubsection{Separation of Telescopes}

The optimal separation between the telescopes was chosen to be 
80\,m.  This
provides the best combination of flux sensitivity and energy
threshold.  Increasing the spacing does not improve flux
sensitivity
in the 200\,GeV - 1\,TeV energy range but it does increase the
energy
threshold.  Decreasing the spacing can reduce the energy threshold
somewhat, but the sensitivity is significantly reduced over the
entire
energy range because the stereoscopic capabilities become less
efficient and the angular resolution degrades leading to a dramatic
increase in the number of unrejected background events.

\subsubsection{Optical Design}

The chosen FOV of the telescope is 3.5$^\circ$.  From
experience operating the Whipple telescope, this is the minimum
required for individual telescopes to reconstruct accurately images
from point sources and it also gives significant sensitivity to
off-axis or extended sources.  Increasing the FOV beyond
this will improve the high energy response of the telescope, but
also
increases the energy threshold for the fixed number of channels. 
The chosen FOV is a compromise between these competing priorities.

The focal length chosen for the telescopes is 12\,m, 
which makes them
$f$/1.2 systems for an aperture of 10 m.  Increasing the focal
length from the $f$/0.7 of the Whipple 10\,m
significantly improves the optical quality of the telescopes
by
reducing the effects of aberrations.  This improvement is necessary
to
match the angular size of the PMTs in the VERITAS cameras
($\sim$
0.15$^\circ$) while retaining the 3.5 $^\circ$ FOV.  However,
increasing the focal length beyond 12\,m does
not improve the performance of VERITAS enough to justify the
increased
cost and complexity needed to achieve the required alignment
precision
and stability for the pixel size chosen.  Also, the reduction in
energy threshold for increasing the focal length beyond 12\,m is
outweighed by the less efficient light collection caused by
increasing
the pixel separation to maintain the required field of view.  Thus,
12\,m provides the best combination of improved optical quality,
sufficient field of view and small pixel spacing.

\subsubsection{Trigger}

The telescope trigger requires 3 adjacent PMTs to record more than
4.2
photoelectrons (p.e.) within $\approx$ 10\,ns.  The 4.2\,p.e. trigger
threshold
level is set by the sustainable single telescope trigger rate at
which
VERITAS can operate (1\,MHz) and by the requirement that the array
trigger
rate from fluctuations of the night-sky background (NSB) not exceed
10\% of the total cosmic-ray rate of 1\,kHz.

\subsection{Collection area}

The collection area of the array reflects a combination of the area
over which the array can detect $\gamma$-ray showers and the
efficiency for retaining those events after analysis.  The
effective collection
area of VERITAS as a function of photon energy is shown in
Figure 4
%Figure~\ref{coll-area-fig} 
for the PMT trigger threshold of
4.2\,p.e..  
For comparison, the
effective area of the Whipple telescope near 1\,TeV is
approximately
4$\times$10$^4$\,m$^2$ \cite{Hillas96}.  The effective collection
area at low energies is reduced because these showers emit fewer
Cherenkov photons and so must arrive with optimum geometry to
trigger
the telescope and allow for event reconstruction.  For the highest
energies, the collection area levels out because the limited FOV of
the telescope truncates these images which develop low in the
atmosphere.

[INSERT FIGURE 4]

\subsection{Energy threshold}

The {\it peak} energy of VERITAS is defined as the energy 
at which the
differential rate of {\it reconstructed} $\gamma$-rays from the
Crab
Nebula per unit interval of energy reaches its maximum. 
Reconstructed
$\gamma$-rays are those whose images pass an image cleaning
procedure
(to eliminate pixels whose light is dominated by fluctuations of
the
NSB light), a reconstructed direction cut, and an image shape cut.
This rate curve is shown in Fig. 5
%Figure~\ref{eth-fig} 
for a
4.2\,p.e. three adjacent pixel trigger threshold and indicates an
{\it array peak} energy of
approximately 75\,GeV. This {\it array peak} energy is substantially 
higher than the
{\it trigger peak} energy i.e., the peak in the $\gamma$-ray rate
for
events triggering the telescope, because low energy Cherenkov
events
which trigger the telescope often do not produce enough light to be
used in the analysis.  However, events at energies below the array
peak energy which trigger the array can still be successfully
reconstructed,
as shown by the extent of the rate curves below the peak. Strong
sources with spectral cut-offs below the array peak energy may
still be detectable.

[INSERT FIGURE 5]

A peak energy of 75\,GeV is a factor of 4 lower than that reached
by
existing imaging Cherenkov telescopes.  This improvement comes from
the combination of the increase in mirror area, the stereoscopic
reconstruction of events, which
permits effective background rejection at lower light levels than
a
single telescope, and the multi-telescope trigger, which permits
the
operation of the individual telescopes much further into the region
where fluctuations in the NSB dominates the single telescope
trigger rate.

The conventional definition of the energy threshold is the point
where the differential 
rate of $\gamma$-rays per unit energy interval from the Crab
Nebula is the highest; we prefer to call this the peak energy since
the
actual threshold may be considerably below this level
 (Figure 5).
%(Figure~\ref{eth-fig}).

\subsection{Energy resolution}

The energy resolution of VERITAS will be considerably better than
that of a single atmospheric imaging telescope such as 
the Whipple telescope because (1) the shower core
location will be known with an accuracy of a few meters, and (2)
several telescopes will view each event at
different distances from the shower core. With a single telescope,
the core distance can be measured
using the location of the image centroid relative to the source
position, but the correlation is not very tight and the
corresponding
energy estimate is not very accurate.  For the Whipple telescope,
the root mean square (RMS) energy resolution using this technique
is $\Delta E/E \approx
0.35$ \cite{Mohanty98}.

The overall energy RMS resolution of VERITAS will be 
${\Delta E/E} \approx $ 0.10 -- 0.15
 from 0.2 to 10\,TeV.
The resolution improves slowly as the energy of the shower
increases.
Below 200\,GeV, the energy resolution degrades.  This
low energy resolution will likely be improved through more
sophisticated energy estimates and through the use of more
restrictive
cuts on events used in the spectral analysis.

\subsection{Angular resolution}

The angular resolution is defined as the width of a two-dimensional
Gaussian fit to the distribution of reconstructed directions for
individual photons from a point source \cite{weekesveritas}.  The
angular resolution of
VERITAS for a single photon as a function of peak energy is
shown in Figure 6. 
It is convenient to characterize integral parameters of the array, such as
angular resolution or flux sensitivity, as functions of array peak energy.
 The majority of events reconstructed with a
given angular resolution will correspond to gamma rays with energies equal to
the peak energy.
%Figure~\ref{ang-res-fig}.  
VERITAS will have better angular resolution
than any existing detector (in space or on the ground) operating
above
a few MeV.  The main reason for the excellent angular resolution of
VERITAS is its stereoscopic imaging capability.  For a single
telescope, the photon arrival direction is well-defined only in the
direction perpendicular to the axis of the image, though the
direction
parallel to the direction of the image axis can be roughly
estimated
\cite{Buckley98}.  Multiple sampling of the shower from the
VERITAS
telescopes overcomes this difficulty and, for larger showers, can
average out fluctuations.  The 
improvement in angular resolution alone leads to
better background rejection than existing single telescopes and it
also will allow the emission regions of extended sources to be
mapped
to arc-minute accuracy. 
In the absence of test beam, the angular 
resolution will be verified by observations of a point source,
e.g., AGN.

[INSERT FIGURE 6]

\subsection{Integral Flux sensitivity}

The performance of VERITAS is summarized by its 
$\gamma$-ray flux sensitivity
versus energy.  The minimum detectable flux of $\gamma$-rays is
defined by the confidence level required for detection or by the
statistics of the detected photons.  A 5$\sigma$ excess of
$\gamma$-rays above the background, or 10 photons (below this,
Poisson
statistics must be used) is required.  The flux sensitivity is
estimated for 50
hours of observations on an object with a differential spectrum given by
dN/dE$\propto$E$^{-2.5}$, as seen from the Crab Nebula in this
energy
range.  Direction and image shape cuts are applied to reject
background events and optimize sensitivity.

This conservative estimate of the
 $\gamma$-ray integral flux sensitivity of VERITAS for point
sources as a
function of array energy threshold is shown in Figure 7.
%Figure~\ref{sens-fig}.
The complex shape of the sensitivity curve is caused by different
energy regions being dominated by the different backgrounds shown.
%Figure~\ref{sens-fig}.  
For energies above 2 -- 3\,TeV, the
sensitivity of VERITAS is limited by photon statistics.  Larger
telescope FOVs can improve this sensitivity in the
future,
as can large zenith angle observations.  In the region near 1\,TeV,
the sensitivity is limited by those rare cosmic-ray protons which
mimic $\gamma$-rays by converting most of their energy into an
electromagnetic cascade in the first few interactions (prompt decay
of
$\pi^0$).  In the energy region between 200\,GeV and $\sim$1\,TeV,
the
background rejection of VERITAS is so good that diffuse cosmic-ray
electrons are expected to dominate the background instead of
hadronic cosmic rays.
The region below 200\,GeV is limited by hadronic cosmic-ray events
whose reconstruction is significantly affected by small NSB
fluctuations.  The lower curve indicates a
relatively dark observation region (near the zenith with no bright
stars in the FOV) while the upper curve indicates a region
where the NSB light is approximately 4 times brighter (as in some
regions of the Galactic plane).

[INSERT FIGURE 7]

In Table~\ref{lab-table} the performance
characteristics of VERITAS are summarized; these are optimized for
point source detection.

\begin{table}
%\begin{center}
\caption{VERITAS Sensitivity}
\label{lab-table} 
\begin{tabular}{lll} 
\hline\hline
Characteristic & E & Value \\ \hline
Peak Energy$^a$ & & 75 GeV \\
Flux sensitivity$^b$ & $>$100 GeV &
9.1$\times$10$^{-12}$cm$^{-2}$s$^{-1}$ = 15\,mCrab \\
 & $>$300\,GeV & 8.0$\times$10$^{-13}$cm$^{-2}$s$^{-1}$ = 5\,mCrab
\\
 & $>$1\,TeV & 1.3$\times$10$^{-13}$cm$^{-2}$s$^{-1}$ = 7\,mCrab \\
Angular resolution & 50\,GeV & 0.14$^\circ$ \\
 & 100\,GeV & 0.09$^\circ$ \\
 & 300\,GeV & 0.05$^\circ$ \\
 & 1\,TeV & 0.03$^\circ$ \\
Effective area & 50 GeV & 1.0$\times$10$^3$m$^2$ \\
 & 100\,GeV & 1.0$\times$10$^4$m$^2$ \\
 & 300\,GeV & 4.0$\times$10$^4$m$^2$ \\
 & 1\,TeV & 1.0$\times$10$^5$m$^2$ \\
Crab Nebula $\gamma$-ray rates& $>$100\,GeV & 50/minute \\
 & $>$300\,GeV & 7/minute \\
 & $>$1\,TeV & 1/minute \\
Energy resolution$^c$  & & $<$15\% \\ \hline
\multicolumn{3}{p{3.9in}}{\small$^a$Energy at which the rate
of photons per unit energy interval from the Crab Nebula is highest
for
a 4.2 photoelectron trigger threshold.} \\
\multicolumn{3}{p{3.9in}}{\small$^b$Minimum integral flux for 
a 5$\sigma$ excess (or $\ge$10 events) in 
50 hours of observations of a source with a Crab Nebula-like spectrum.} \\
\multicolumn{3}{p{3.9in}}{\small$^c$RMS ($\Delta E/E$)} \\
\end{tabular}
%\end{center}
\end{table}

%%%%% Technical Description
%%%%%%%%%%%%%%%%%%%%%%%%%%%%%%%%%%%%%%%%%%%%%%%%%%

\section{Technical Description}
%FINAL VERSION
\subsection{Telescope Design}

The Davies-Cotton optical design \cite{Davies57} used in the
Whipple 10\,m
reflector (and in many subsequent $\gamma$-ray telescopes) has 
mirror facets
 which are spherical and identical, facilitating fabrication
at a reasonable cost and making alignment easy. This design 
also has smaller
off-axis aberrations than a parabolic reflector so that it has good
image quality out to a few degrees from the optic axis.  Its one
limitation, that the surface is not isochronous, means that the
reflector introduces a time-spread into the light pulse. 
Alternative designs were investigated but it was concluded
that the Davies-Cotton collector provides the
optimal combination of optical quality and cost
effectiveness. Increasing the $f$-number from $f$/0.7 to $f$/1.2
substantially improves the optical quality of the telescope
(Figure 8).
%(Figure~\ref{images}). 
It is then possible to match the inherent
fluctuations in the shower images with a camera that has a
reasonable number of pixels ($\approx$ 500) and appropriate FOV
($\approx$3.5$^\circ$). The larger f/number reduces the time spread
to $<$ 4 ns (which is a good match to the inherent width of the
shower pulse). 

[INSERT FIGURE 8]

The desirable optical parameters of the telescope follow from the
characteristics of a $\gamma$-ray shower \Cv image.  For a
$\gamma$-ray source at the center of the FOV, the
average image centroid (distance) is $\approx$ 0.85$^\circ$ from the
center of the focal plane, and the centroid 
moves to slightly larger values at
larger energies. The image has a RMS width and length of about
0.14$^\circ$ and 0.25$^\circ$, respectively. The reflector will
have sufficient resolution to record image structure on this scale. 

A key factor in the reflector quality and cost is the mirror
facets. Although a number of mirror materials and fabrication
technologies have been developed and investigated for VERITAS, a
conventional approach has been chosen as the most cost effective. 
The
facets will be made of float glass, ground or slumped, polished,
aluminized and anodized. They will be hexagonal in shape to permit
closepacking and will be 60\,cm across the flats.

\subsection{Mounts: Optical Support Structure/Positioner}

The VERITAS telescopes will duplicate the Whipple telescope in
using
a custom-designed, welded-steel, space-frame 
Optical Support Structure (OSS) mounted
on 
a commercial positioner. They will have a f/number of 1.2,
a 
rectangular OSS to facilitate the use of
mirror covers, and tighter performance specifications than the Whipple
10\,m reflector. 

Simpson, Gumpertz and Heger (SG\&H), the firm that designed 
the Whipple 10 m telescope, have conducted two studies and
a 
conceptual design.  Cost and performance were evaluated as the 
critical parameters of aperture, f-number, wind loading, storm 
survival, stiffness, and pointing were varied.  Based on these 
studies, SG\&H determined that a trussed-steel OSS of 10 x 11 m 
aperture and 12 m focal length (f/1.2) could be designed to match 
commercially available positioners (Figure 9).
%(Figure \ref{telescope}). 
Although initially the OSS will have 250 mirrors (collection 
area = 75\,m$^2$), it will be capable of supporting 315
mirrors (collection area = 100\,m$^2$).
SG\&H confirmed that a f/1.2 OSS
can be constructed locally from standard materials to meet VERITAS 
design specifications.

[INSERT FIGURE 9 HERE]

The telescope RMS blur, RMS decentering, and RMS pointing 
specifications, 
with and without corrections (determined by offset-facet alignment and/or
lookup tables),
are 0.02/0.04 $^\circ$, 0.01/0.05 $^\circ$, and 0.02/0.05 
$^\circ$, respectively for wind speeds up to 20 m.p.h. 
Azimuth slew speed will be 
1\,$^\circ$/s and elevation slew speed will be 
0.5\,$^\circ$/s. 
The total weight of the OSS, mirrors, counterweights and the
detector is estimated to be 16,000 kg. Commercially available
positioners 
from radio and radar antenna manufacturers can support this weight 
and meet the pointing specifications. 

When not in use, the mirrors on the reflector will be covered by
four ``window-blind'' covers which will be mounted on the sides
of the OSS.

\subsection{Electronic Camera}

\subsubsection{Outline}

The electronic camera for each VERITAS telescope is similar to
previous imaging atmospheric \Cv detectors \cite{Cawley90} but with
some significant improvements. An outline of the major components of the
camera is shown in Figure 10.
%Figure \ref{efig1}.  

The array trigger is formed from a
combination of individual telescope triggers at the central
trigger. The trigger is designed to record all $\gamma$-ray events
that contain enough light to be usefully identified as $\gamma$-ray
events (reconstruction threshold) while limiting the trigger rate
to a manageable level.

When a readout trigger
decision has been made for an individual telescope, a sub-array of
telescopes or the full array, the control electronics loads the
appropriate section of the FADC memory of those signal channels
with hits onto a data bus which is read out by the controller and
stored in reflective memory modules.  This information is read out
by local CPUs and merged into event data for the telescope.  The
individual
telescope data is transmitted to the central station, merged with
data
from the other telescopes and stored. A centrally generated optical
pulse is used to synchronize individual electronic channels and the
time tagging of events across the array.

[INSERT FIGURE 10]

Some of the individual elements of the VERITAS electronic cameras
are discussed briefly below.

\subsubsection{Front-end Electronics}

\paragraph{Photomultiplier Tubes:}

A low noise, high gain ($>$ 10$^6$), photon counting detector, with
fast time response (risetimes $<$ 2.5 ns) matched to the intrinsic
temporal width of the $\gamma$-ray induced air shower is necessary
to increase the signal-to-noise ratio and to operate VERITAS at the
lowest possible energy.  These requirements are presently
satisfied only by PMTs. The spacing between the centers of
the PMTs corresponds to a focal
plane angular distance of $0.15^\circ$. The active light collection
diameter of the PMTs is 25\,mm.  The collection efficiency is
increased by the use of light concentrators. Tests have
identified the Hamamatsu R7065, Electron Tube 9142W,
 and the Phillips Photonis XP2900 as PMTs which
satisfy the VERITAS requirements.

\paragraph{High Voltage Supply:}

Modular commercial high voltage supplies will be used so that each
PMT has a separately programmable high voltage. These are used to
adjust the PMT gains and to reduce the current draw when a bright
star image
falls directly on a specific PMT. A suitable unit which
provides a 900 volt swing in 48 channel modules is available
from CAEN. The high voltage
is supplied by
cables attached to each PMT from a remote system crate.  Commands
to each telescope system are sent from the central station via
Ethernet.

\paragraph{Signal Amplifiers:}

A custom circuit with a linear amplifier is attached to each PMT in
the focus box. This circuit is based on a standard integrated
circuit amplifier chip, e.g., National Electronics Semiconductors CLC449 
which have a bandwidth of $\sim$1\,GHz. The targeted gain of
the PMT and amplifier provides an output threshold signal
level of $\sim$2\,mV/photoelectron and a dynamic range of 1,800.
The amplifier also provides a secondary input to the data
acquisition system which permits charge injection for calibration
and diagnostic purposes during daytime.

\paragraph{Cabling:}

Pulses are transmitted from the focal plane PMTs to the telescope
electronics through 40 m of high quality coaxial cable. This
introduces some time dispersion into the signals. An analog optical
fiber system with sufficient dynamic range
(1:1000) is being tested on the Whipple 10\,m telescope
\cite{Bond00}.  Fibers introduce no dispersion and are
light-weight. If these tests are satisfactory optical fibers would
be an attractive upgrade alternative to coaxial cabling.

\subsubsection{Trigger Electronics}

VERITAS must trigger at low PMT threshold levels 
to achieve its low
energy threshold, and with simple trigger schemes this would naturally
lead to very high trigger rates. To limit the overall trigger rate
and to keep the data acquisition deadtime to an acceptable level,
a multi-level trigger scheme 
has been developed.  This system keeps the overall trigger rate below
 1 kHz. The  four trigger levels are defined as:

\begin{description}
\itemsep=0pt

\item[{\rm \underline{Level 1 (CFD)}}] 
The outputs from the PMTs come directly to constant fraction
discriminators (CFDs). The analog bandwidth of this system is
$>500$\,MHz.
The discriminator threshold and output pulse width are programmable
by command.  This module
also has a short adjustable delay element to allow the timing in
each channel to be matched to better than 2\,ns.  An analog fanout
of the PMT signal is provided from this board to the FADC system.
The CFDs constitute the Level 1 trigger. The NSB 
produces PMT singles rates 
given by the dotted curve in Figure 11. 
%Figure \ref{efig3}. 
At low thresholds
the dominant contribution to the trigger rate comes from
fluctuations in
the number of single photoelectrons (p.e.) produced by 
NSB. At higher thresholds the background rate is dominated by PMT
afterpulsing. At a threshold of 4.2\,p.e., the Level 1 trigger rate
is $\sim$1\,MHz for each channel.

\item[{\rm \underline{Level 2 (PST)}}]  A hardware pattern trigger
at each telescope based on Level 1 triggers reduces the
background by discriminating between photon-initiated shower
images, which are
compact, and background triggers caused by sky noise or
afterpulsing,
which have random locations in the camera plane. The dashed curve
in Figure 11 
%Figure \ref{efig3} 
shows the expected rate of these background
events when the coincidence requirement is that at least three nearest
neighbors in the 499 pixel camera are hit. The pattern selection
trigger (PST) follows the scheme presently being used at the
Whipple Telescope \cite{Bradbury99}.

\item[{\rm \underline{Level 3 (Array)}}] The central station
receives 
the Level 2 triggers from each telescope. The Level 3 trigger
system selectively delays each Level 2 signal to account for the
wavefront orientation and determines if the overall array trigger
condition is satisfied.  The trigger condition depends on the
array configuration and the observing strategy, but typically
requires the time coincidence of several telescopes.  The solid
curve in Figure 11
%Figure~\ref{efig3} 
shows the expected background rates for
this trigger if three of seven telescopes fire within 40\,ns coincidence
time.  If the Level 3 trigger condition is met, readout of the
telescope event information is initiated.

\item[{\rm \underline{Level 4}}] The background trigger rate can be
reduced by demanding that the individual telescope trigger clusters
conform to the predicted parallactic displacement of the
$\gamma$-ray images.  Thus, a high level trigger, implemented
either in hardware or in software, can be used to significantly
suppress the hadronic background.  This system is currently in the
development stage.

\end{description}

[INSERT FIGURE 11]

\subsubsection{Signal Recording}

\paragraph{Flash ADCs:}

The FADC provides a digitized version of the \Cv pulse waveform,
giving
the maximum information possible about the shape and time
structure.
These devices allow operation of the individual telescopes well
into
the NSB, eliminate the need for delay cables
(minimizing signal dispersion), and allow real-time calibration of
the
PMT and signal cable propagation times.  Also, the time structure
of
the pulse is expected to become wider near the edge of the \Cv
light
pool, and with the FADCs these effects can be corrected for in
later
analysis of the digitized pulses.  The increasing availability of
high
bandwidth FADC chips for commercial purposes makes it possible to
include a FADC in every detector channel of VERITAS. 
Commercial modules are expensive and VERITAS will use custom-built
units. A prototype FADC system has been developed and tested on the
Whipple telescope and appears to have all the necessary parameters
for VERITAS.
 Each FADC channel uses a commercially available 8-bit FADC
integrated circuit. This system has a sample rate of 500\,MHz,
a memory depth of 64 microseconds, and a novel autoranging gain switch
that provides a dynamic range of 0 to 1020 with no loss of signal
bandwidth \cite{Buckley99}.

\paragraph{Data Acquisition:}

The data acquisition electronics are based largely around the VME
standard
electronic architecture: a fast VME backplane and distributed
computation performed by local CPUs running a real-time operating
system. Each controller will be connected, using fast, fiber-optic
connections to a local workstation which in turn is connected to a
central high speed switch connected to the central CPU. The central
CPU will perform control and quicklook functions, further data
compression and integration of the distributed data. This system
will operate using cyclic and multi-thread processes, adhering to
POSIX standards where possible in order to maintain future systems
compatibility.  Secondary systems to control mount movement and
high voltage generation will be Pentium PCs running a UNIX-like
operating system such as Linux.  Communication will be via
Ethernet connection.  For a telescope readout rate of $\sim$1\,kHz,
each telescope is expected to have an average data flow rate of
$\sim$1,600\,kbyte/s, resulting in a 400\,kbyte/s  
rate
on any VME backplane.  This rate includes real \Cv events and
spurious background hits. The VERITAS data acquisition system will
operate well within this limiting range.

\subsection{Calibration}

     The seven telescope VERITAS array has substantially better
energy resolution ($\Delta E/E \approx 15\%$) than the single
Whipple 10m telescope ($\approx 35\%$), thereby requiring a
substantially improved calibration system. The maintenance of the 
larger number of data acquisition channels requires an automated
system for both optical and electronic injection into 
the PMTs and preamplifiers. The widely-spaced array also
requires special attention to details such as nightly calibration
of telescope gains and multi-telescope event timing
synchronization. 

	The angular resolution cannot be measured directly
in the absence of test beam, but it 
can be verified by observations of a point source,
e.g. an AGN.
 
The VERITAS calibration system has been designed to provide
automated, redundant measurements of all relevant
calibration parameters as well as to test system
functionality automatically during the daytime. It consists of three different
subsystems: optical injection, electronic or charge injection, and
atmospheric monitoring. The optical injection system consists of a
central pulse light source with variable optical attenuation which
is distributed to all telescopes through uv-transmitting fiber
optic cables. The optical pulses are used to 
illuminate the camera directly and to measure the mirror reflectivity
by bouncing the light off the mirror surface before recording the
optical pulse with the camera. The charge injection system injects
a computer-adjustable pulse height and pulse width into each
preamplifier input in order to measure electronic gain, as well as
to verify each channel's operation.  The charge injection system is
located inside the camera, and is capable of triggering any
possible combination of tubes by setting an internal `mask'. 
This capability allows one to generate automatically various tube
hit configurations to diagnose triggering capability of all four
trigger levels. The atmospheric monitoring system consists of
several radio-controlled  light sources on the side of Mt. Hopkins
as well as a Polaris monitor/star tracker telescope. Both the light
sources and the star tracker/Polaris monitor use CCD cameras with
UBVRI filter wheels to spectroscopically analyze the recorded light
sources. The overall absolute calibration uncertainty for the
combined calibration system will be approximately 15\%, appropriate
for the expected energy resolution of the VERITAS array.

%%%%% Scientific Motivation
%%%%%%%%%%%%%%%%%%%%%%%%%%%%%%%%%%%%%%%%%%%%%%%%%%

\section{Scientific Objectives}
%FINAL VERSION

\subsection{Active Galactic Nuclei}

 The AGN detected at $\gamma$-ray energies are blazars which are
generally radio-loud, flat-spectrum sources 
and radio and X-ray selected BL Lacertae (BL Lac) objects.   
These sources are characterized by highly variable
and predominantly non-thermal emission.
Currently, no model for the production of $\gamma$-rays in the 
AGN is generally
accepted.  The two most popular models are those in which high
energy
electrons (known to be present in the jets from radio, optical, and
X-ray observations) produce the $\gamma$-rays by inverse Compton
scattering of low energy photons and those in which very high
energy
protons produce the $\gamma$-rays by initiating cascades within the
jets.

VHE observations of the blazars Markarian 421 (Mrk 421)
and Markarian 501 (Mrk 501) have revealed
extremely variable VHE emission (Fig.12)
%(Figure~\ref{main-m597-fig};
\cite{Quinn98,Gaidos96}.  Whipple and EGRET observations
have also led to reformulations of blazar unification models based
on
intrinsic source luminosity rather than solely on the effects of
orientation differences \cite{Ghisellini99,Georganopoulos98}.

[INSERT FIGURE 12]

Despite these gains, there is much to learn about these enigmatic
objects.  VERITAS can help answer many of the unknowns concerning
blazars, particularly when combined with lower energy observations
by
GLAST, X-ray and optical telescopes.  The study of
AGN will be approached from two perspectives: detailed studies
involving 
observations of individual objects to reveal the physics operating
in
the AGN, and broad studies where more VHE sources are detected to
illuminate the properties of different object classes.

\subsubsection{Detailed studies:}

The large effective area of VERITAS enables accurate measurements
of
extremely rapid variations in the $\gamma$-ray flux as illustrated
in Figure 13.
%Figure~\ref{main-960515-fig}.  
The left part of the figure shows
Whipple observations of the fastest flare ever recorded at
$\gamma$-ray energies.  Although the flare was clearly detected, the
structure of the flare may not have been fully resolved.  The dashed curve in
the
figure is a hypothetical flux variation which matches the Whipple
data.  The right part of the figure shows a simulation of what
VERITAS
would detect above 200\,GeV.  All features of the flare are clearly
resolved.  VERITAS can be divided into two dedicated sub-arrays of
3-4
telescopes, with one sub-array observing a single object for as
long
as it is above the 30$^\circ$ observation ``horizon'' for as many
nights as a multi-wavelength campaign lasts.  Thus, VERITAS can
provide more complete measurements of flaring activity and greatly
increase the chance of seeing a wide range of activity in the
objects
studied.  The sensitivity of VERITAS to short timescale variations
will be an excellent match to X-ray and optical telescopes,
permitting
much improved measurements of energy-dependent variability.

\subsubsection{Broad studies:} 

The flux sensitivity of VERITAS will permit the detection
of
weaker sources of VHE emission.  Its sensitivity at low energies will
permit the viewing of objects further from Earth (the optical depth
for pair production decreases rapidly with decreasing energy) and
those objects which have spectral cut-offs below the sensitive
range
of existing Cherenkov telescopes.  Based on the multi-wavelength
spectra of Mrk 421 and Mrk 501 and the Spectral
Energy Distributions of known BL Lacs,
VERITAS will detect 30 or more X-ray selected BL Lacs.

[INSERT FIGURE 13]

EGRET sources associated with AGN now number at least 67
\cite{Hartman99} and span a range of redshifts from z = 0.03 
to z = 2.28. However
only six AGN have been detected above 300\,GeV.  The
most likely causes of this disparity are intrinsic cut-offs in the
source
spectra and, for the more distant sources, attenuation of the TeV
$\gamma$-rays by pair-production with background IR radiation.  The
low energy threshold and good sensitivity of VERITAS will overcome
these hurdles and lead to the detection of approximately 15 of the
EGRET AGN, more if GLAST or all-sky X-ray monitors indicate when
these
objects are in high emission states.  VERITAS is in a unique
position
to make these measurements, providing better photon statistics than
space-based telescopes at these energies and superior energy
resolution and sensitivity relative to other ground-based
telescopes.

\subsection{Shell-type supernova remnants}

SNRs are widely believed to be the sources of hadronic cosmic rays
up
to energies of approximately $Z\times 10^{14}$\,eV, where $Z$ is
the
nuclear charge of the particle. However, a clear indication for the
acceleration of hadronic particles in SNRs is still missing.  The
existence of energetic {\it electrons} is well-known from
observations
of synchrotron emission at radio and X-ray energies (e.g.,
\cite{Koyama95}).  Recently, the detection of TeV $\gamma$-rays
from
the shell-type remnants, SN\,1006, RXJ\,1713.7-3946 and Cassiopeia\,A
has
been reported \cite{Weekes99}.  When combined with X-ray and radio
observations, the VHE $\gamma$-ray observations provide a means
of
resolving the various contributions (proton-proton collisions or
electron bremsstrahlung or inverse Compton scattering). The
combined
observations can also give information about SNR shell environments
such as the maximum particle energy and magnetic field.  Both 
quantities are
important, but unknown, parameters in shock acceleration theories.

Figure 14 
%Figure~\ref{main-snrlims-fig}
 shows the EGRET measurements and
Whipple
upper limits for the SNR IC\,443.  These results already eliminate
much of the allowed parameter space for $\gamma$-ray emission from
these objects (from hadrons and electrons), and begin to raise some
questions about the validity of current models for the objects
studied
and  even for the SNR origin of cosmic rays.

[INSERT FIGURE 14]

For a typical SNR luminosity and angular extent, VERITAS should be
able to detect objects within 4\,kpc of Earth according to one
popular
model of $\gamma$-ray production by hadronic interactions
\cite{Drury94}.  Approximately, twenty shell-type SNRs with known
distances lie within this distance range, permitting VERITAS to
investigate which characteristics of SNR are necessary for them to
be
sites of particle acceleration.

\subsection{Diffuse galactic $\gamma$-ray emission:}

High energy $\gamma$-rays traverse the Galaxy without significant
attenuation implying that the diffuse emission probes high energy
processes in the Galaxy as a whole.  EGRET studies \cite{Hunter97}
show generally good agreement with detailed models in both 
spatial and spectral features.  A striking exception is that
there is a 40\% excess in measured flux
at the highest EGRET energies, 
and the measured spectrum systematically rises above predictions. 
In some models, e.g., \cite{pohl98,moskalenko00} this is
attributed to inverse Compton scattering from SNR 
electrons injected with a hard spectral index, although this appears
to be in contradiction to upper limits at TeV
energies \cite{lebohec00}.    Detailed spectral studies in the 
GeV/TeV energy bands are necessary to understand the discrepancies.

\subsection{Compact Galactic Objects}
\label{compact-gal-sect}

\subsubsection{Pulsar-powered nebulae:}

With the flux sensitivity of VERITAS 
it will be possible to detect
Crab Nebula-like objects anywhere within the Galaxy if their declination
is
$>-28^\circ$. The energy resolution and broader energy
coverage will significantly improve spectral measurements leading
to accurate estimates of the nebular magnetic field and maximum
electron energy, and to more detailed tests of $\gamma$-ray emission
models.  With its excellent angular resolution, VERITAS may even 
resolve the GeV-TeV emission region of nearby objects like
the
Crab Nebula.

\subsubsection{Gamma-ray pulsars:}

Attenuation of $\gamma$-rays by pair-production interactions in the
intense magnetic field near pulsars leads to a super-exponential
cut-off in the
spectra predicted by polar cap models.  Because the outer gap
models
do not predict such sharp cut-offs, the detection of pulsed GeV-TeV 
$\gamma$-rays may be decisive in favoring the outer gap model over
the
polar cap model. The excellent sensitivity and energy resolution of
VERITAS will allow spectral measurements to be conducted even 
though these objects are expected to have rapidly falling spectra in
the GeV-TeV range.

\subsubsection{Unidentified galactic EGRET sources:}

A legacy of EGRET is 170 unidentified sources, many of which are
in the Galactic plane \cite{Hartman99}.  The positional uncertainty
of these sources make identifications with sources at longer
wavelengths unlikely.  VERITAS will have the sensitivity and low
energy threshold necessary to detect many of these objects if their
spectra do not cut off above 10 GeV.  Detailed
studies with the exellent source location capability of VERITAS
could
lead to identifications with objects at longer wavelengths.

\subsubsection{Galactic plane survey:}

Because of its flux sensitivity and large field of view, one of the
first priorities of VERITAS will be to conduct a sensitive survey
in
the 100\,GeV -- 10\,TeV energy range.  To maximize the scientific
returns, the first task will be to survey the region near the
Galactic
plane.  VERITAS can observe the plane over the longitude range
$0^\circ < l < 245^\circ$, but the initial survey will concentrate
on
the denser regions.  For an 80-night survey of the region $0^\circ
< l
< 85^\circ$, VERITAS will be sensitive to fluxes down to $\sim$0.02
Crab, above 300\,GeV.

\subsection{Cosmic Ray Composition}

VERITAS, with its fine pixellation, large mirror area, and
 multi-telescope image measurement can measure the Cherenkov light
emitted by the primary cosmic ray nucleus before it interacts with
the atmosphere, thereby providing a high-resolution 
($\Delta$Z/Z $<$ 5\% for Z $>$ 10) 
charge measurement of cosmic rays
around the knee of the  all-particle spectrum \cite{kieda01}. 
This measurement will be essentially independent of any assumed
nuclear interaction model;  therefore VERITAS can provide a tagged
nuclear beam that will determine the true air-nucleus 
interaction characteristics as a function of primary charge and
energy.  This should  provide a method to eliminate interaction
model-dependent biases in the interpretation of these results,
 as well as allow other experiments to use the experimentally
determined interaction model to
 reanalyze their data.  In addition, VERITAS will be sensitive to  
nuclei heavier than iron in the
PeV energy region, to exotic particle states
such as relativistic magnetic monopoles \cite{wick01} or strange
quark matter \cite{banerjee99}.

\subsection{Background Infrared Radiation}

In traversing intergalactic distances, $\gamma$-rays can be
absorbed
by photon-photon pair production on background photon fields
\cite{Gould67}.  For VHE photons, the most important interaction
is
with infrared (IR) fields.  Measurements of VHE spectra from
extragalactic sources permit indirect investigation into the nature
of
the IR background field because distortions in the $\gamma$-ray
spectrum due to pair-production depend on the density spectrum of
the
background field.  VHE spectra from Mrk 421 and Mrk 501 have
already
been used to set upper limits on the IR background from 0.025\,eV
to
0.3\,eV \cite{Stecker98,Dwek92,Biller98a,vassiliev01}.  At some
wavelengths, these limits are as much as an order of magnitude
below
the upper limits set by the DIRBE/COBE satellite (see Fig. 15).
%Figure~\ref{main-irlim-fig}).  
The current limits on the IR density
are $\sim$5 to 10 times higher than predicted from galaxy 
formation and evolution
(e.g., \cite{Primack99,Malkan98}).  However, with better 
spectral measurements of
existing sources and detections of more objects, particularly at
higher redshift, VERITAS can substantially improve these limits,
and
may eventually detect the effects of the IR field.

[INSERT FIGURE 15]

\subsection{Gamma-ray Bursts}

The sources and mechanism for producing Gamma-ray Bursts (GRBs)
 remain
unknown.  The detection by EGRET of an 18\,GeV photon from a GRB more than
90
minutes after the burst was detected by BATSE \cite{Hurley94}
demonstrates that high energy $\gamma$-rays play an important role
in
the energetics of GRBs.  Also, because the high energy emission can
be
delayed, it can be pursued with rapid follow-up observations.  More
recently, the Milagro Collaboration has reported a weak ($\sim
3\sigma$) detection of TeV emission from a GRB \cite{McEnery99}
raising the possibility that TeV photons may carry a significant
fraction of the energy emitted in a GRB.  Because of its low energy
threshold, VERITAS will be able to see bursts with hard spectra
out to z$\sim$1 or
more.

\subsection{Neutralino annihilation in the Galactic center}

Current astrophysical data would appear
to indicate the need for a cold dark matter
component with $\Omega \approx 0.3$ (see \cite{Gawiser98,
Primack98,Kamionkowski98}).  A good candidate for
this component is the neutralino, the lightest stable
supersymmetric
particle.  If neutralinos do comprise the dark matter and are
concentrated near very massive astrophysical objects like
the center of the Galaxy, their direct
annihilation
to $\gamma$-rays should produce a unique signal not easily mimicked
by
other astrophysical processes: a monoenergetic annihilation line
with
mean energy equal to the neutralino mass.  Cosmological constraints
and limits from accelerator experiments restrict the neutralino
mass
to the range 30\,GeV - 3\,TeV.  Thus, VERITAS and GLAST together
will
allow a sensitive search over the entire allowed neutralino mass
range.  Recent estimates of the annihilation line flux for
neutralinos
at the galactic center \cite{Bergstrom98} using a
galactic model with central cusps in the density distribution of
the
dark matter halos \cite{Navarro96} predict a
$\gamma$-ray signal which may be of sufficient intensity to be
detected with VERITAS and GLAST.  The
better sensitivity and lower energy threshold of VERITAS will be
critical to covering a broad part of the allowed range of
neutralino
and dark matter parameters.

\subsection{Quantum gravity}

Quantum gravity can manifest itself as an effective
energy-dependence
to the velocity of light in a vacuum. This dependence is
 caused by propagation through a
gravitational medium containing quantum fluctuations on distance
scales near the Planck length ($\simeq 10^{-33}$\,cm)
(e.g., \cite{Amelino98}).  

If the quantum gravity correction to vacuum refractive index
exists, it should appear at the energy scale comparable
to Planck mass ($\approx 10^{19}$ GeV). Recent work within the
context of string theory indicates, however, that the quantum
gravity scale may occur at a much lower energy, perhaps
as low as 10$^{16}$ GeV \cite{Witten96}.
TeV observations of variable emission from
astrophysical objects provide a means of searching for the effects
of
quantum gravity \cite{Biller98b}.  VERITAS will significantly
improve short timescale variability measurements and also will
detect more distant objects.  Variability on short timescales from
sources at z$>$0.1 would be sensitive
to quantum gravity effects 
and would provide a test of the validity of Lorentz
symmetry at energies within a factor of five of the Planck mass
(in some models).

\subsection{Primordial black holes}

Primordial black holes, if they exist, should emit a burst of
radiation in the final stages of their evaporation (e.g.,
\cite{Page76}).  In the standard model of particle physics, 
this last burst of radiation should release about 10$^{30}$erg in
1\,s
with the energy distribution peaked near 1\,TeV \cite{Halzen91}. 
In
two years of operation VERITAS would be able to reach a sensitivity
level for this type of evaporation of 700 pc$^{-3}$ yr$^{-1}$. 
More extreme
models (e.g.,
\cite{Hagedorn68}) would produce lower energy events on much
shorter
time-scales which may be detectable with VERITAS using a special
trigger and the flash ADC system \cite{Krennrich99b}.

%%%%% History and Status of Technique
%%%%%%%%%%%%%%%%%%%%%%%%%%%%%%%%%%%%%%%%

\section{Status of VHE Gamma Ray Astronomy}
There
are seven groups now using the atmospheric 
Cherenkov imaging technique 
\cite{weekes00}. These techniques are relatively mature
and the results from simulaneous 
observations, with different telescopes of the
same source are consistent \cite{Protheroe97}.
Vigorous observing programs are now in place at all of these
facilities.  An important milestone for the field has been reached
in
that both galactic and extragalactic sources have been reliably
detected \cite{Weekes99}.

To exploit fully the potential of ground-based $\gamma$-ray
astronomy
the detection techniques must be improved. This will happen by
extending the energy coverage of the technique and by increasing
its
flux sensitivity. Ideally one would like to do both, but in practice
there must be trade-offs.  Reduced energy thresholds can be achieved
by
the use of larger, but cruder, mirrors. This approach is
currently
being exploited using existing arrays of solar heliostats: STACEE
\cite{Chantell98}, CELESTE \cite{Giebels98}, Solar-2
\cite{zweerink99}, GRAAL \cite{plaga99}. 
These projects may achieve thresholds as low as
20-30\,GeV where they will effectively fill the current gap in the
$\gamma$-ray spectrum from 20 to 200\,GeV but with poor energy
resolution and small fields of view. 
A European project
(MAGIC) \cite{lorenz99} to build a 17\,m aperture telescope is now
under way and a similar project, MACE has been proposed in India
\cite{Bhat00}. 
Ultimately this gap will be
covered by GLAST with weaker point source sensitivity at the higher
energies.  This next generation $\gamma$-ray space telescope is
scheduled for launch in 2005 by an international collaboration
\cite{Gehrels99}. Extension to higher energies ($>$10\,TeV) can be
achieved by atmospheric Cherenkov telescopes working at large
zenith
angles and by particle arrays at very high altitudes. The MILAGRO
water Cherenkov detector in New Mexico \cite{Sinnis95} operates 24
hours a day with a large field of view and has good sensitivity to
$\gamma$-ray bursts and transients.

The primary objective of VERITAS will be to have high sensitivity
in
the 100\,GeV to 10\,TeV range. The German-French HESS (initially
four
and eventually perhaps sixteen 12\,m class telescopes) will be
built
in Namibia \cite{Konopelko99,hofmann99} and the Japanese
CANGAROO-III array
(with four telescopes in Australia) \cite{Matsubara97,mori99} will have
similar objectives for observations in the southern hemisphere.
MAGIC, HESS and CANGAROO are approved projects with target
completion
dates in 2002 to 2004. The arrays will exploit the high sensitivity
of
the atmospheric Cherenkov imaging technique and the high
selectivity
of the array approach.  The relative flux sensitivities as a
function
of energy are shown in Figure 16,
%Figure~\ref{main-senscomp-fig}, 
where the
sensitivities of the wide field detectors are for one year and the
ACT
are for 50 hours; in all cases a 5\,$\sigma$ point source detection
is required. 
The flux sensitivity of HEGRA, CAT and CANGAROO-II are very
similar to that of Whipple; similarily the flux sensitivity of HESS and
CANGAROO-III are anticipated to be similar to that of the simulated
sensitivity of VERITAS. Hence the integral flux sensitivities shown
in Figure 16 for Whipple and VERITAS are representative 
of the other observatories. 

[INSERT FIGURE 16]

It is apparent from this figure that, on the low energy side,
VERITAS
will complement the GLAST mission and will overlap with STACEE and
CELESTE. At its highest energy it will overlap with the Tibet Air
Shower Array \cite{Amenomori97}. It will cover the same energy
range
as MILAGRO but with greater flux sensitivity.  As a northern
hemisphere telescope VERITAS will complement the coverage of
neutrino
sources to be discovered by AMANDA and ICE CUBE at the South Pole.

\section{Acknowledgements}
This work was supported in part by grants from the U.S. Department of
Energy,
the National Science Foundation, the Smithsonian
Institution, PPARC in the U.K. and
Enterprise-Ireland in Ireland. The referees are thanked for helpful
suggestions.

%%%%%%%%%%%%%%%%%%%%%%%%%%%%%%%%%%%%%%%%%%%%%%%%%%%%%%%%%%%%%%%%%
%%%%%%%%%%%%%
%\input{vp-fig}

%Figure 1.
%\newpage
\begin{figure}
\centerline{\epsfig{file=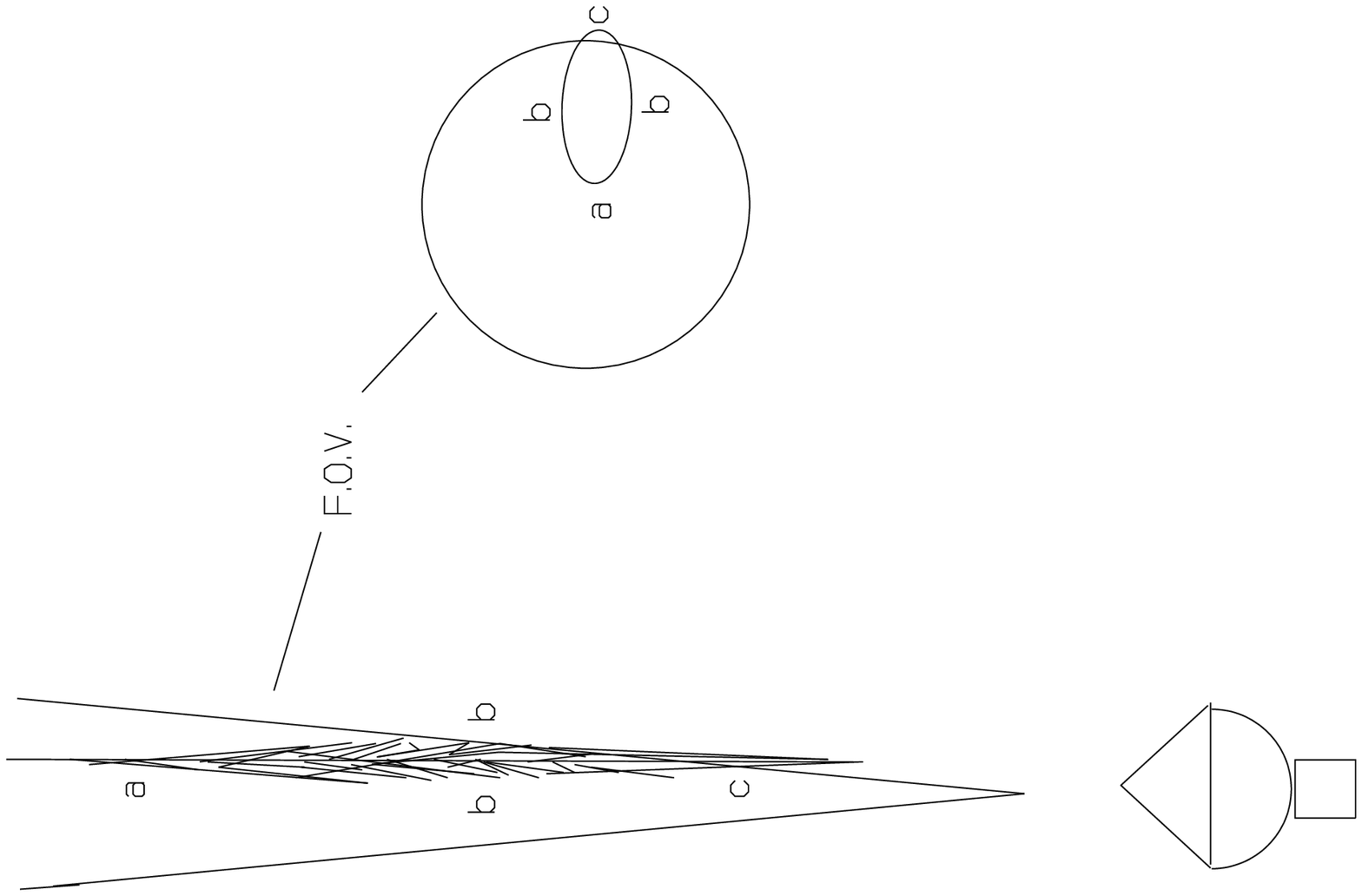,width=5.0in,angle=270.}}
\vspace{2cm}
\caption{The imaging technique. The cone of
acceptance of the camera intercepts the core of the air
shower. The elliptical contour of a typical Cherenkov light image
as
seen in the focal plane of the camera (typically of diameter
3.5$^\circ$) as seen looking into the camera is shown on the
 right; note the left to right inversion. Images of $\gamma$-ray showers
coming from a source parallel to the optic axis are narrow and
point
towards the center. Images from background cosmic rays are broader
and
have no preferred pointing direction.
\label{shower}
}
\end{figure}

%Figure 2.
\newpage
\begin{figure}
%\vspace*{-0.35in}
\centerline{\epsfig{file=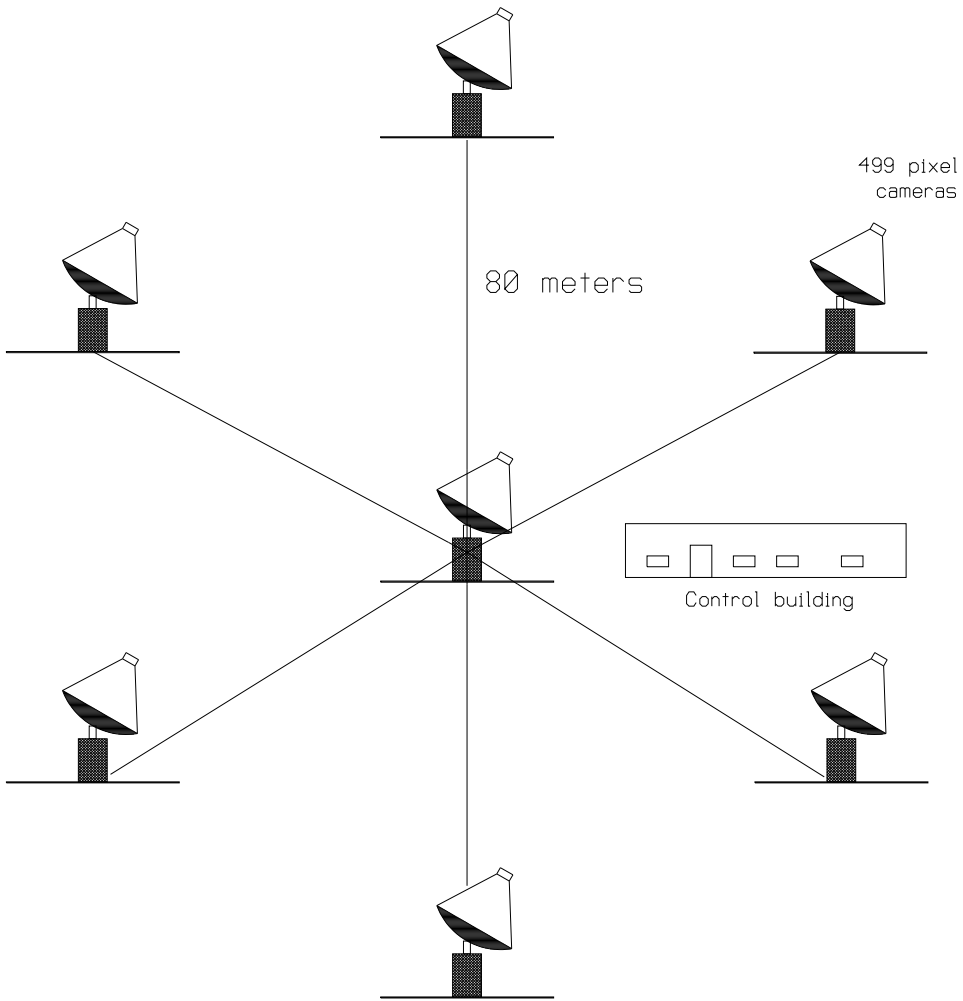,height=5.05in,angle=270.}}
\vspace{2cm}
\caption{Layout of telescopes in VERITAS. 
\label{athena-fig}
}
\end{figure}

%Figure 3
\newpage
\begin{figure}
\centerline{\epsfig{file=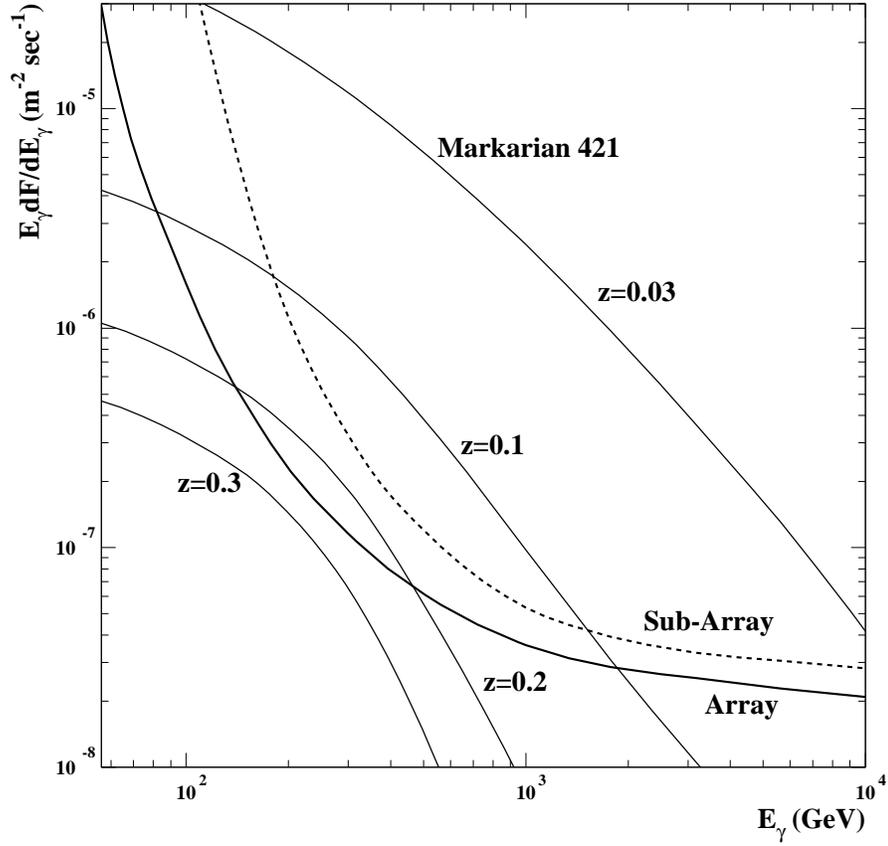,height=5in}}
\vspace{5cm}
\caption{The differential flux sensitivity of full array and sub-array
(three telescopes) of VERITAS is shown as a function of energy. The
measured spectrum of Mrk 421 
in flaring state is shown (z = 0.03); 
the spectra of the
 same source if it were at z = 0.1, 0.2 and 0.3 are also shown.
\label{fluxsensitivity}}
\end{figure}

% Figure 4
\newpage
\begin{figure}
\centerline{\epsfig{file=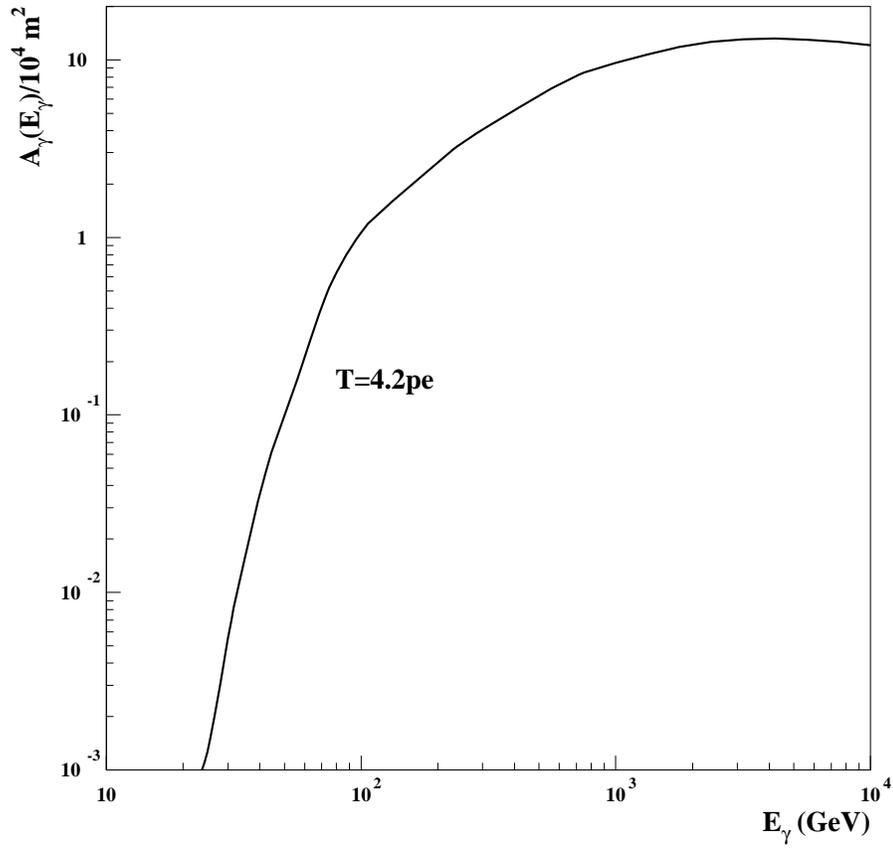,width=5.0in}}
\vspace{2cm}
\caption{The collection area of VERITAS as a function of 
gamma-ray energy.  
\label{coll-area-fig}
}
\end{figure}

%Figure 5
\newpage
\begin{figure}
\centerline{\epsfig{file=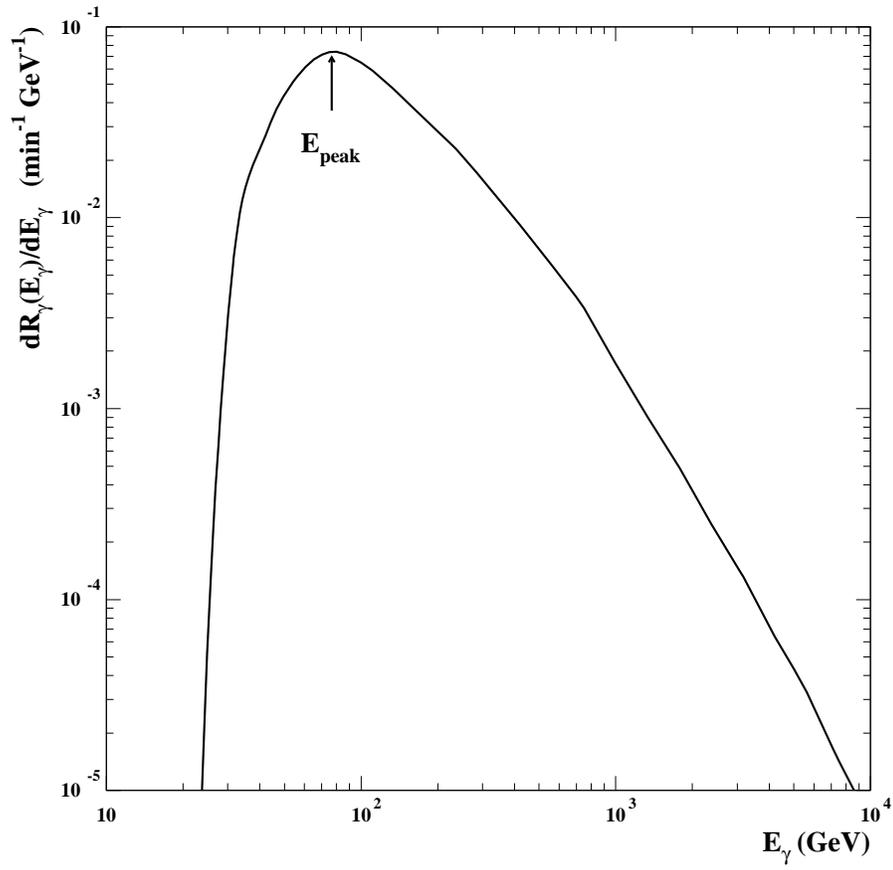,width=5.0in}}
\vspace{2cm}
\caption{Differential detection rates of the Crab Nebula for
VERITAS when the telescope trigger is from three adjacent pixels.  
The peak of the curve is sometimes referred to as the energy
threshold (see text) but here is called the peak energy.  
\label{eth-fig}
}
\end{figure}

%Figure 6
\newpage
\begin{figure}

\centerline{\epsfig{file=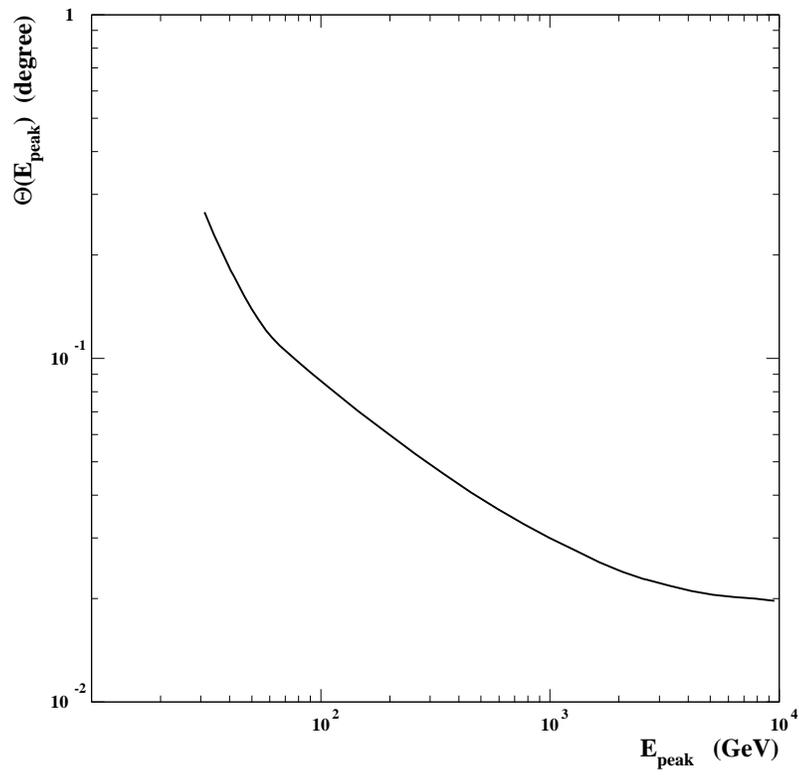,width=4.5in}}
\vspace{2cm}
\caption{Angular resolution of VERITAS for a single photon
as a function of array $\gamma$-ray peak energy.  
\label{ang-res-fig}
}
\end{figure}

%Figure 7
\newpage
\begin{figure}
\centerline{\epsfig{file=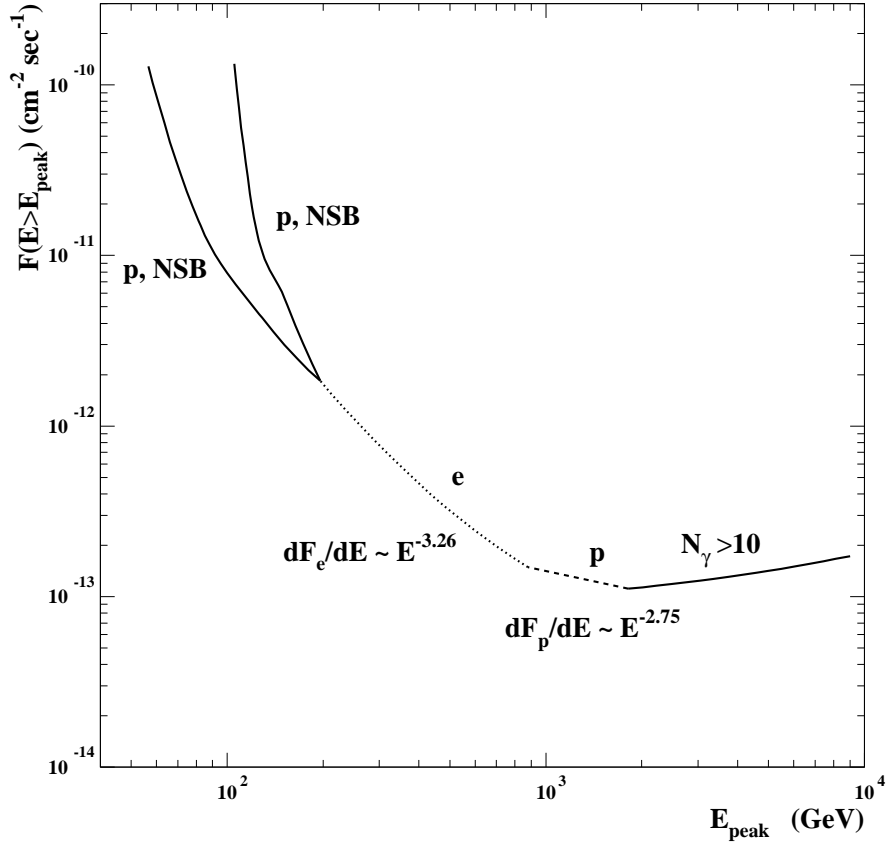,width=5.0in}}
\vspace{2cm}
\caption{The $\gamma$-ray sensitivity of VERITAS to 
point-like sources in 50 hours of observing.  The dominant 
background as a function of energy
threshold is indicated (see text for details).  The two curves at
low energies indicate the sensitivity of VERITAS in dark (lower curve)
and bright (upper curve) night-sky background (NSB) regions. P 
refers to the hadronic background, e to the electron background.
A minimum of ten photons is required at high energies. These
are the sensitivities at small zenith angles; at large zenith
angles the sensitivity improves for higher energies.
\label{sens-fig}}
\end{figure}

%Figure 8
\newpage
\begin{figure}
\centerline{\epsfig{file=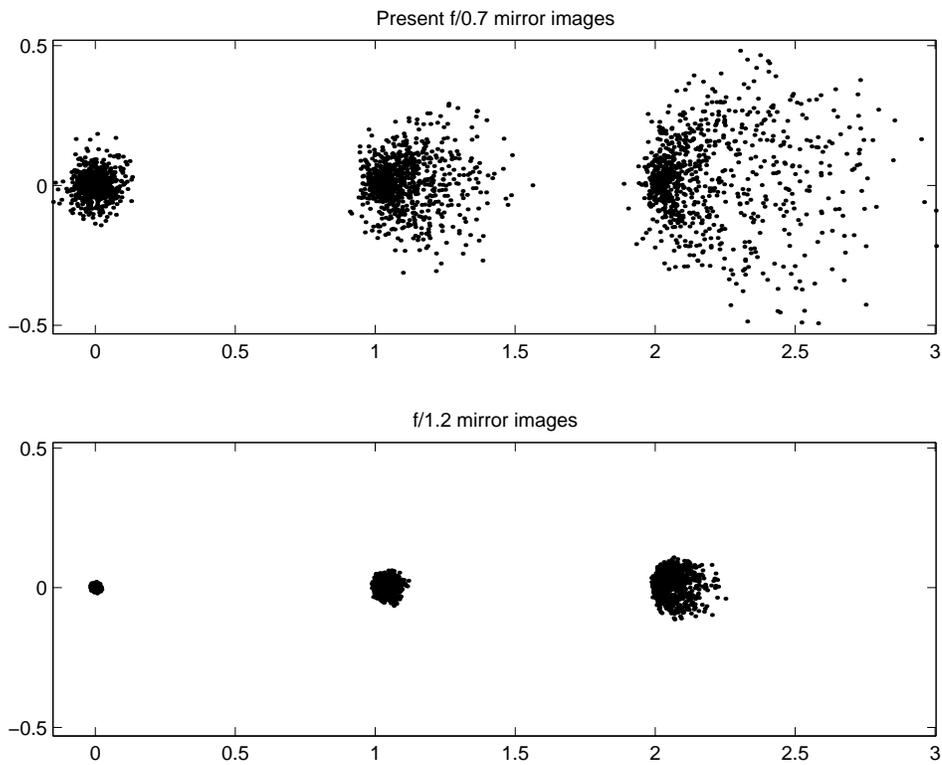,width=5.0in}}
\vspace{2cm}
\caption{Point spread images in the 
focal plane corresponding to a source at 0$^\circ$, 1$^\circ$, and
2$^\circ$  off axis for $f$/numbers of 0.7 (top, Whipple
telescope) 
and 1.2 (bottom, VERITAS).  The facet size is 60 cm and the
image
size is given in degrees.  A total of 1000 points lie in each
image.
\label{images}
} 
\end{figure}

%Figure 9
\newpage
\begin{figure}
\centerline{\epsfig{file=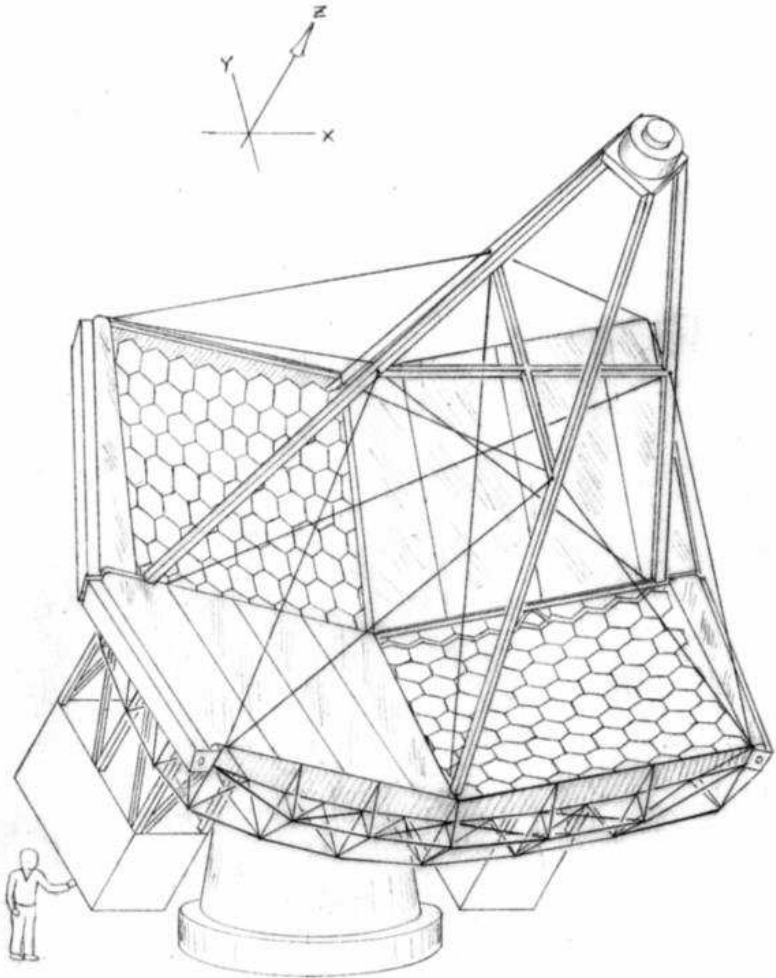,height=5.5in}}
\vspace{2cm}
\caption{Engineering sketch of VERITAS telescope based
on conceptual design with covers retracted on upper left and lower right
quadrants. The telescopes will not initially be populated by mirrors
in corners. Sketch by T.Hoffman.
\label{telescope}}
\end{figure}

%Figure 10
\newpage
\begin{figure}
\centerline{\epsfig{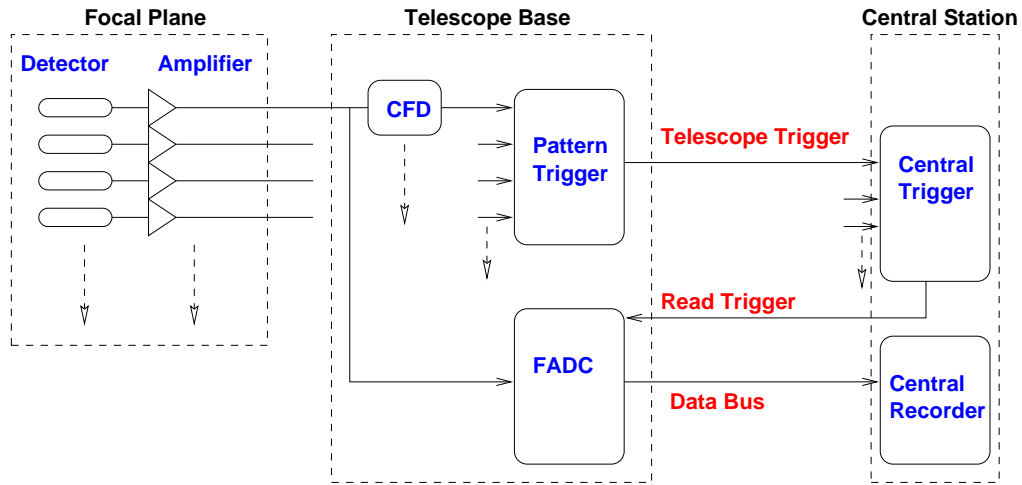}}
\vspace{2cm}
\caption{Outline of electronics. The PMT signals are 
amplified in the camera head and transmitted to the electronics
located in a building near the telescope base. The input analog
signals are split with one output fed to fast constant fraction
discriminators (CFDs) to form the initial trigger by detecting
coincidences on a time-scale from 5 to 15\,ns.  The outputs from
these discriminators are fed to a local
pattern trigger unit which tests for adjacent detector hits in the
telescope. The other branch of the analog signal from each detector
is fed to a flash analog-to-digital converter (FADC) which
digitizes the detector waveform into a circulating memory.  
This results in a digitized version of the signal pulse from each
pixel. The telescope trigger signal is sent to the central trigger
location where programmable delays are applied to the trigger
signal from each telescope to correct for the difference in the
time of arrival of the  wavefront at the individual telescopes
due to the source location in the sky.
\label{efig1}}
\end{figure}

%Figure 11
\newpage
\begin{figure}
%\vspace*{-0.7in}
\centerline{\epsfig{file= 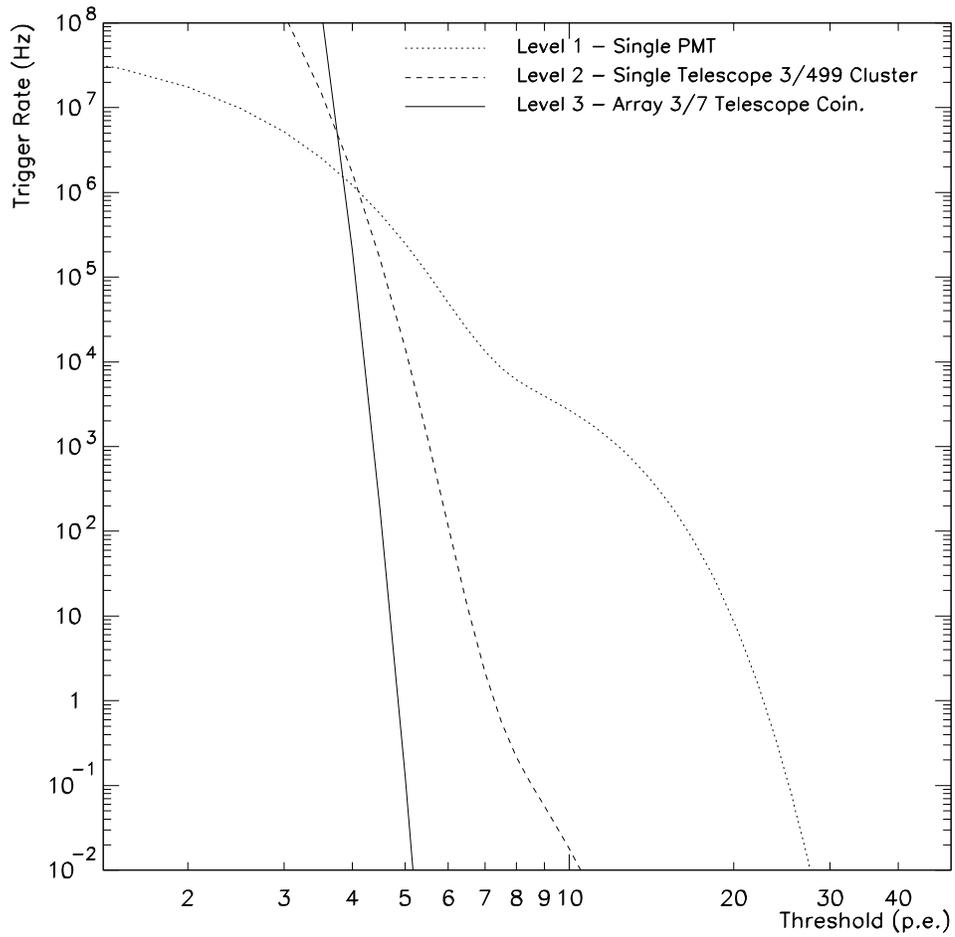,width=5.0in}}
\vspace{2cm}
\caption{Expected rates for the various trigger levels of  VERITAS
as a function of PMT threshold. The various curves are
discussed in the text.
\label{efig3}
}
\end{figure}

%Figure 12
\newpage
\begin{figure}
\centerline{\epsfig{file=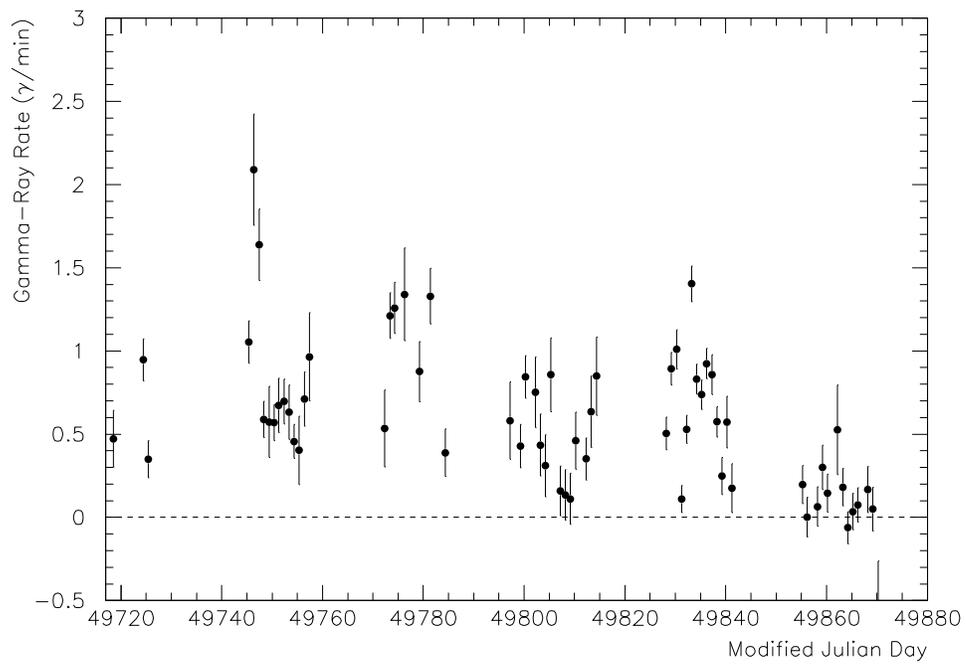,width=5.0in}}
\vspace{2cm}
\caption{Variability in $\gamma$-ray signal from Markarian 501 in
1997.}
\label{main-m597-fig}
\end{figure}

%Figure 13
\newpage
\begin{figure}
%\centerline{\epsfig{file=main_m4_960515_col.ps,width=5.0in}}
\centerline{\epsfig{file=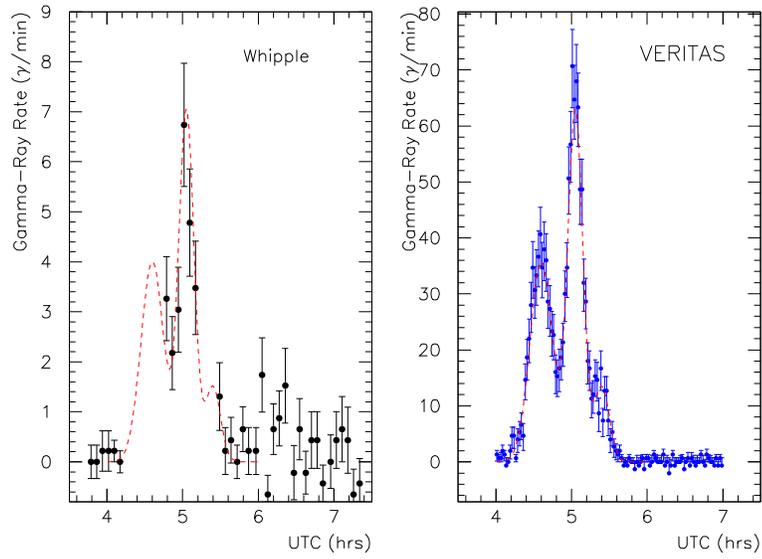,width=5.0in}}
\vspace{2cm}
\caption{{\it Left:} Whipple observations of a rapid flare from Mrk
421 on 1996 May 15.  The superimposed
dashed
curve is a possible intrinsic flux variation which is consistent
with
the VHE data.  {\it Right:} Simulated response of VERITAS to such
a
flare above 200\,GeV.
\label{main-960515-fig}
}
\end{figure}

%Figure 14
\newpage
\begin{figure}
\centerline{\epsfig{file=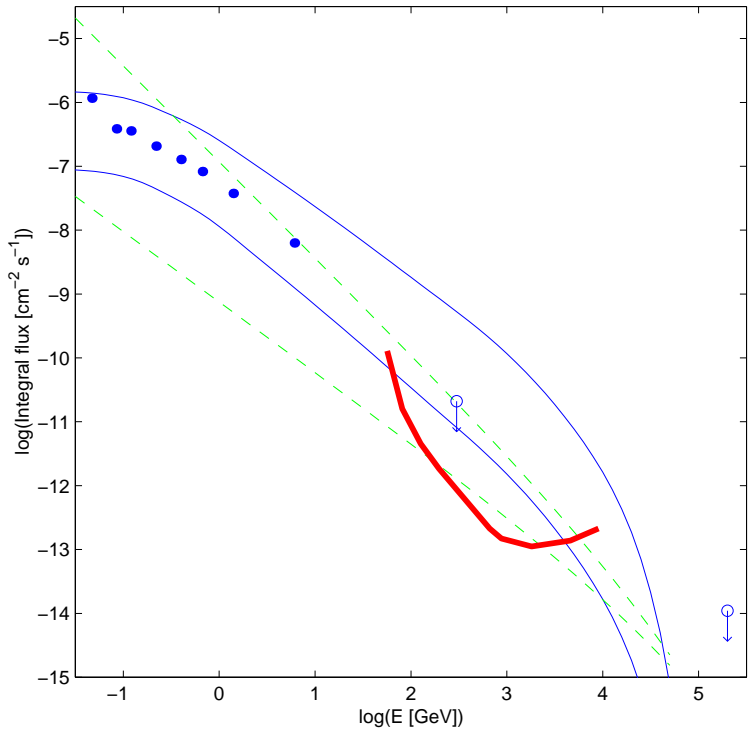,width=4.5in}}
\vspace*{0.5in}
\caption{Predicted $\gamma$-ray spectra in the shell-type SNR
IC\,443.  The region between the solid lines depicts the spectra
from
hadronic interactions for a range of model parameters. 
  The region between the dashed lines is the
predicted inverse-Compton spectra for $B$ between 20\,$\mu$G and
50\,$\mu$G and the range of flux normalizations and electron
spectra allowed by the X-ray observations.  EGRET data points
(filled
circles) and the current upper limits (open circles) from Whipple
and the
Cygnus air shower array are shown.  The sensitivity of VERITAS for
a
50 hour observation of this object is indicated by the thick curve.
\label{main-snrlims-fig}
}
\end{figure}
%\newpage

%Figure 15
\newpage
\begin{figure}
\centerline{\epsfig{file=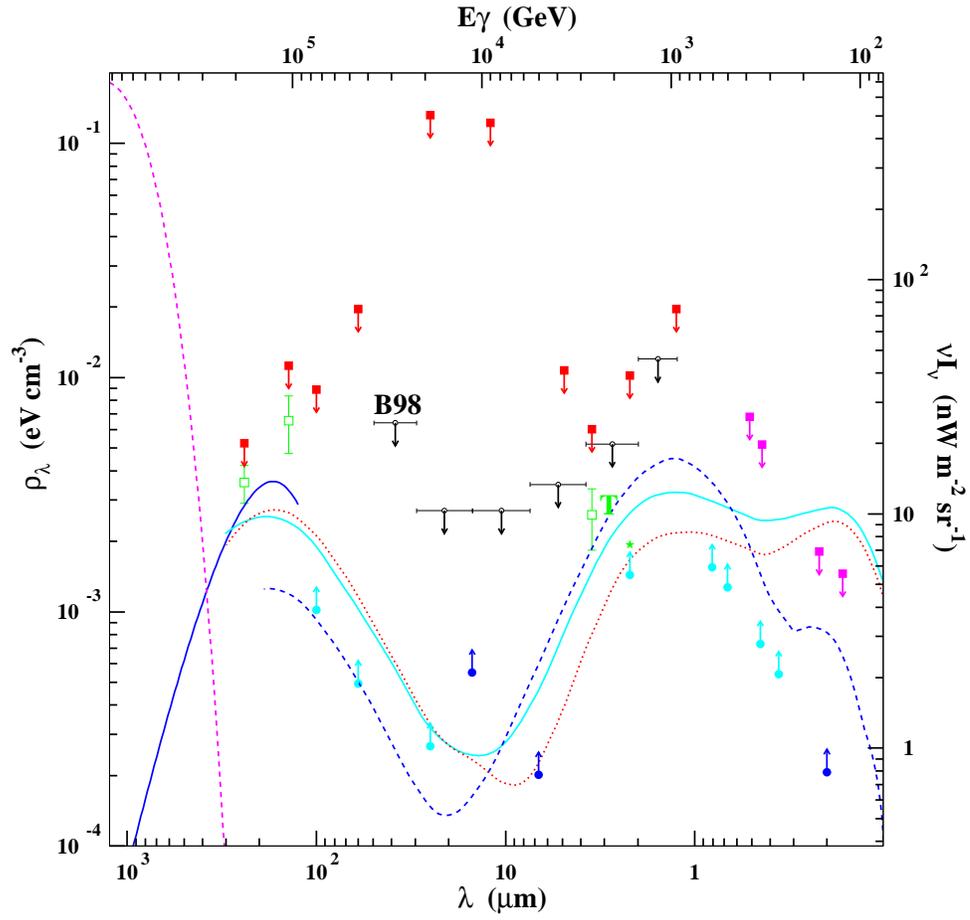, width=5in}}
\vspace{2cm}
\caption{Limits on the Extragalactic IR Background derived from DIRBE
and other experiments. The limits from TeV measurements are shown as
broad arrows while the predictions of theoretical models are shown
as dotted lines.}
\label{main-irlim-fig}
\end{figure}

%Figure 16
\newpage
\begin{figure}
%\vspace*{-0.7in}
\centerline{\epsfig{file=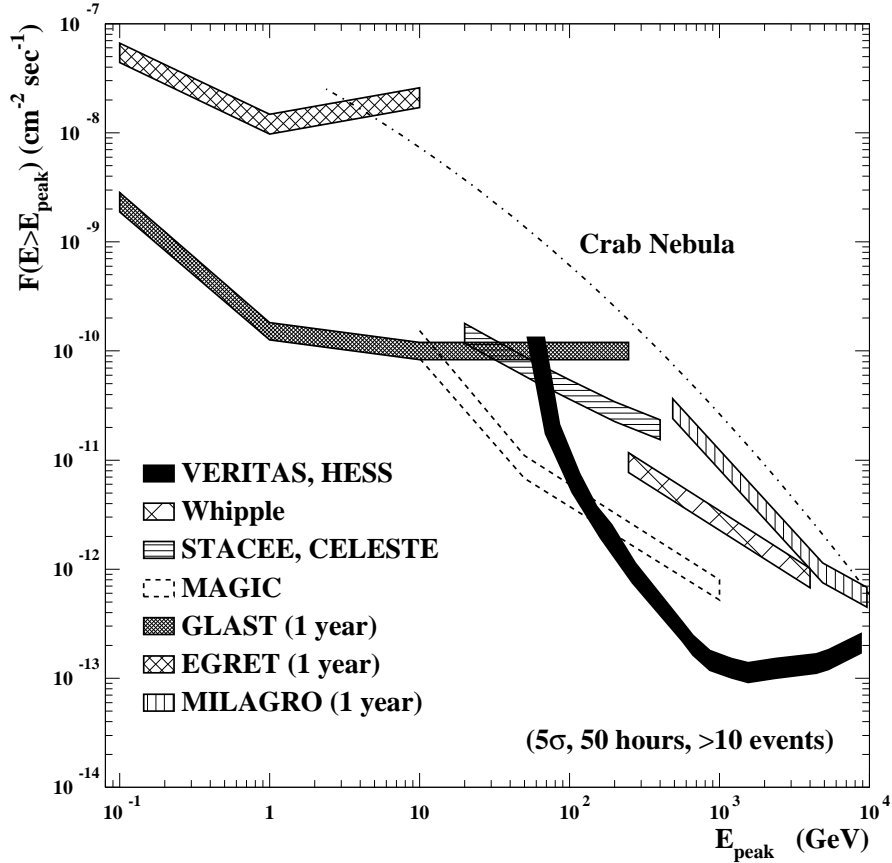,width=5in,angle=0.}}
\vspace{2cm}
\caption{Comparison of the point source sensitivity of VERITAS to
Whipple, MAGIC, CELESTE/STACEE; GLAST, EGRET
and MILAGRO. The sensitivity of MAGIC is
based on the availability of new technologies, e.g., high quantum
efficiency PMTs, not assumed in the other experiments. EGRET, GLAST
and MILAGRO are wide field instruments and therefore ideally suited
for all sky surveys.  The turn-up in the VERITAS sensitivity at
higher
energies is primarily caused by the limited field of view of the
cameras and the requirement of VERITAS detecting at least 10
photons.
\label{main-senscomp-fig}
}
\end{figure}

\end{document}